\documentclass[a4paper,11pt]{article}
\usepackage{pos}
\usepackage{bbold}
\usepackage{graphicx} 
\usepackage{booktabs} 
\usepackage{slashed}
\usepackage{physics}
\usepackage{simpler-wick}
\usepackage[absolute,overlay]{textpos}
\title{Strange and charm contributions to nucleon charges and moments\footnote{
The work of RZ is supported by the US National Science Foundation under grant PHY 1653405 ``CAREER: Constraining Parton Distribution Functions for New-Physics Searches''. 
We thank MILC Collaboration for sharing the lattices used to perform this study. The LQCD calculations were performed using the Chroma software suite~\cite{Edwards:2004sx} with the multigrid solver algorithm~\cite{Babich:2010qb,Osborn:2010mb}. 
This research used resources of 
the National Energy Research Scientific Computing Center, a DOE Office of Science User Facility supported by the Office of Science of the U.S. Department of Energy under Contract No. DE-AC02-05CH11231 through ERCAP; 
the Extreme Science and Engineering Discovery Environment (XSEDE), which is supported by National Science Foundation grant number ACI-1548562;
and facilities of the USQCD Collaboration, which are funded by the Office of Science of the U.S. Department of Energy, 
Extreme Science and Engineering Discovery Environment (XSEDE), which is supported by National Science Foundation grant number ACI-1548562. }}
\ShortTitle{Strange and charm contributions to nucleon charges and moments}

\author*[a,b]{Rui Zhang}
\author[c]{Tanmoy Bhattacharya}
\author[c]{ Rajan Gupta}
\author[a,b]{Huey-Wen Lin}
\author[c]{Santanu Mondal}
\author[c]{Sungwoo Park}
\author[c]{Boram Yoon}

\affiliation[a]{Department of Physics and Astronomy, Michigan State University, East Lansing, MI 48824}

\affiliation[b]{Department of Computational Mathematics,
  Science and Engineering, Michigan State University, East Lansing, MI 48824}

\affiliation[c]{Computer, Computational, and Statistical Sciences CCS-7, Los Alamos National Laboratory, Los Alamos, NM 87545, USA}

\emailAdd{zhangr60@msu.edu}

\abstract{
We present preliminary results for strange and charm contributions to nucleon charges and moments. The scalar, axial and tensor charges, and unpolarized first moments are calculated using clover-on-HISQ formulation and cover four lattice spacings, $a=\{0.06,0.09,0.12, 0.15\}$~fm, and three pion masses, $M_\pi=\{310,220,130\}$~MeV. The renormalization factors are calculated  nonperturbatively using the RI-sMOM scheme. We carry out a chiral and continuum extrapolation to obtain  physical results.
}

\FullConference{%
 The 38th International Symposium on Lattice Field Theory, LATTICE2021
  26th-30th July, 2021
  Zoom/Gather@Massachusetts Institute of Technology
}

\begin{document}
\maketitle

\section{Introduction}
Nucleon charges and moments are important quantities to study for the elucidation of the nucleon structure. The scalar charge plays an important role in dark matter search, and gives the pion-nucleon sigma term. The axial charge is the contribution of quark spins to the proton spin, and enters in the spin-dependent dark matter cross-section. The tensor charge gives the contribution of the quark Electric Dipole Moment (EDM) to the neutron EDM. The first unpolarized moment gives the momentum fraction of the quark, an important quantity to describe the quark PDF  especially for sea quarks. These quantities for the isovector ($u-d$) case have been calculated with high precision using lattice QCD. As the precision of light quark calculation improves, the contributions from heavier quark flavors, strange and charm, also need to be precisely determined.

All operators for charge and moment calculations are quark bilinears,   $O=\bar{q}\Gamma q$. Specifically, for the charge operators, $\Gamma=\mathbf{1}$ for the scalar charge, $\Gamma=\gamma_i\gamma_5$ for the axial charge,  $\Gamma=\sigma_{ij}$ for the tensor charge, are local operators. On the other hand, $\Gamma=D_4\gamma_4-\tfrac{1}{3}\sum_iD_i\gamma_i$ is a non-local operator for the first unpolarized moment, which gives the momentum fraction carried by the quark.

\begin{figure}
    \centering
    \includegraphics[width=0.3\linewidth]{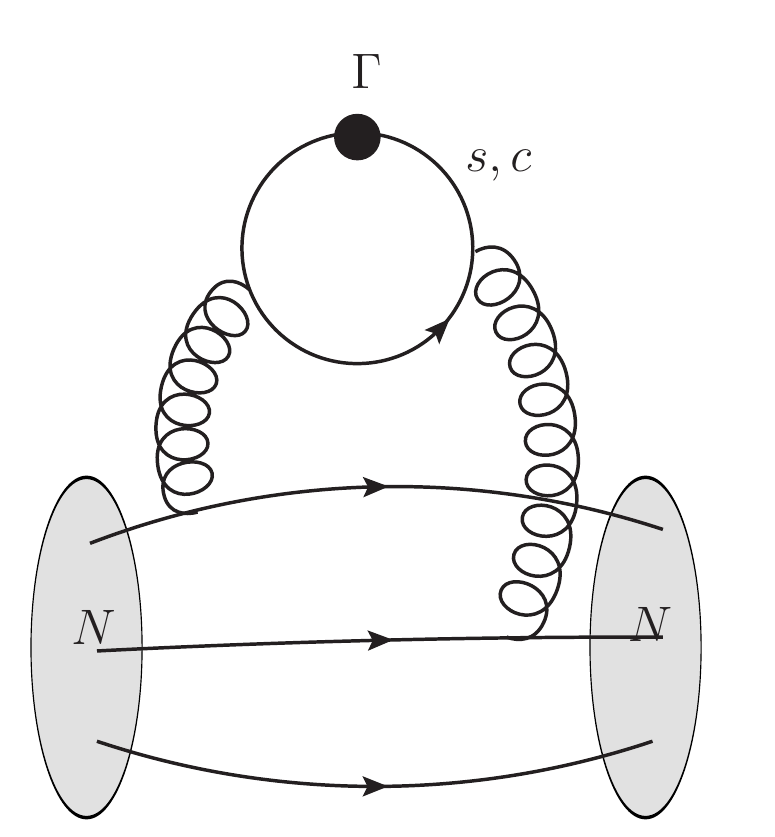}
     \includegraphics[width=0.3\linewidth]{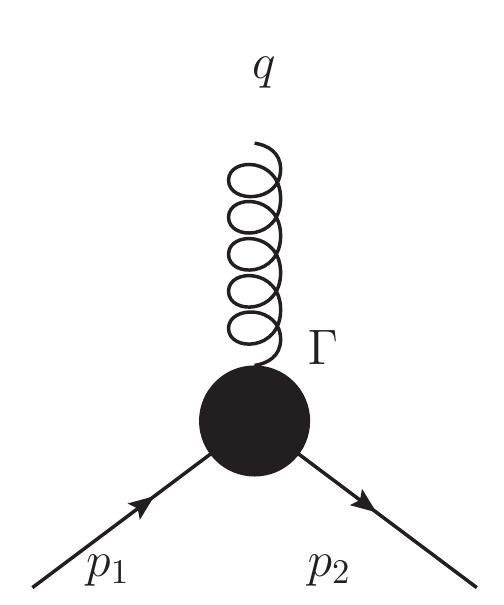}\qquad\qquad
    \includegraphics[width=0.25\linewidth]{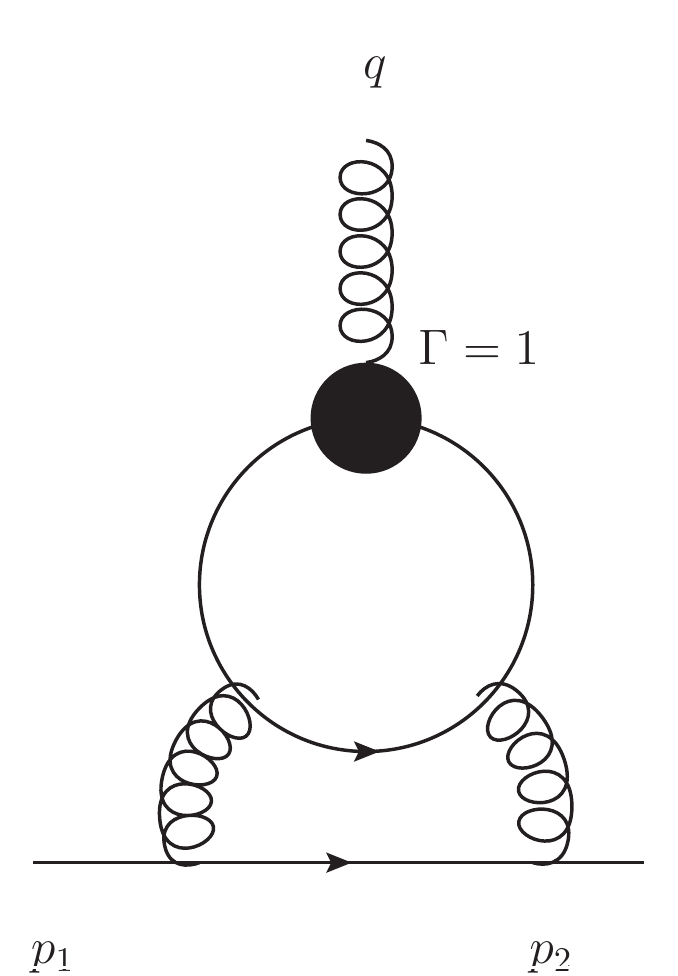}
    \caption{Left: Disconnected diagram on lattice. Connected NPR diagrams (Middle) for all operators and mixing diagrams (right) for the scalar charge. }
    \label{fig:FeymanDiagram}
\end{figure}

For strange and charm quarks, the contribution only comes from disconnected diagrams, ie, the correlation of the self contraction of the quark bilinear operator $\bar{q}\Gamma q$ (forming a quark loop)  with the nucleon, as shown in Fig.~\ref{fig:FeymanDiagram}. The details of disconnected calculations can be found in Ref.~\cite{Bhattacharya:2015wna}.

\section{Lattice Setup, Matrix Elements and Renormalization}
We calculated 2- and 3-point functions on six ensembles generated with 2+1+1 flavors of highly improved staggered quarks (HISQ)~\cite{Follana:2006rc} by the MILC collaboration~\cite{MILC:2012znn}. They cover four lattice spacings, $a\approx\{0.06,0.09,0.12,0.15\}$~fm, and 3 pion masses of $M_\pi\approx \{130, 220, 310\}$~MeV. To reduce statistical noise, the lattices are hypercubic (HYP) smeared~\cite{Hasenfratz:2001hp} before the calculation of the quark propagators. 
The details of the ensembles are summarized in Table~\ref{tab:lat_info}.
 \begin{table}
    \centering
    \begin{tabular}{|ccc|cccc|c|}
    \hline
          a (fm) & $M_\pi$ (MeV) & $L^3\times T$ &  $N_\text{conf}^c$ & $N_\text{conf}^s$ & $N_\text{src}^c$ & $N_\text{src}^s$ &$\frac{N_\text{src}^\text{2pt}}{\text{cfg}}$\\
         \hline
          0.06 & 320(2) & $48^3\times 144$ & 469 & 87 & 4000 & 4000 & 32 \\
         \hline
          0.09 & 313(3) & $32^3\times 96$ & 633 & 889 & 6000 & 6000& 32 \\
         \hline
          0.09 & 138(1) & $64^3\times 96$ & 721 & 310 & 4000 & 4000 & 32 \\
         \hline
          0.12 & 310(3) & $24^3\times 64$ & 983 & 897 & 4000 & 8000 & 32 \\
         \hline
          0.12 & 228(2) & $32^3\times 64$ & 902 & 869 & 5000 & 5000 & 32 \\
         \hline
          0.15& 321(4) & $16^3\times 48$ & n/a & 1795  & n/a & 2000 & 32 \\
         \hline
    \end{tabular}
    \caption{Parameters of the 2+1+1-flavor HISQ ensembles used for the strange and charm contributions to the nucleon charges and moments. $N_\text{conf}$ gives the number of gauge configurations, $N_\text{src}$ is the number of the noise source measurements to approximate the disconnected loop propagators, and $\frac{N_\text{src}^\text{2pt}}{cfg}$ is the number of the two-point sources on each configuration. 
    }
    \label{tab:lat_info}
\end{table}
We computed the nucleon two point correlators of the nucleon interpolating operator \(\chi\) with different projections $\mathcal{P}$:
\begin{equation}
    C^\mathcal{P}_{\text{2pt}}(\tau) = \langle 0| \mathcal{P} \chi(\tau)\bar\chi(0)|0\rangle,
\end{equation}
where $\mathcal{P}^\text{unp}=(1+\gamma_4)/2$ for unpolarized projection, and $\mathcal{P}^\text{pol}=(1+\gamma_4)(1+i\gamma_5\gamma_i)/2$ for polarized projection. Then we compute the disconnected loops:
\begin{equation}
    C^{O}_\text{loop}=\sum_n\Tr\left[\hat{O} S(n)\right]
\end{equation}
These two are combined together to obtain the disconnected contribution.
\begin{align*}
    C^\text{O}_{\text{3pt}}(t,\tau) =&\left\langle\left( C^{\mathcal{P}\phantom{O}}_{\text{2pt}}(\tau)-\langle C^{\mathcal{P}\phantom{O}}_{\text{2pt}}(\tau)\rangle\right)\left(C^O_\text{loop}(t)-\langle C^O_\text{loop}(t)\rangle\right)\right\rangle
\end{align*}

\begin{figure}
    \centering
    \includegraphics[width=0.45\linewidth]{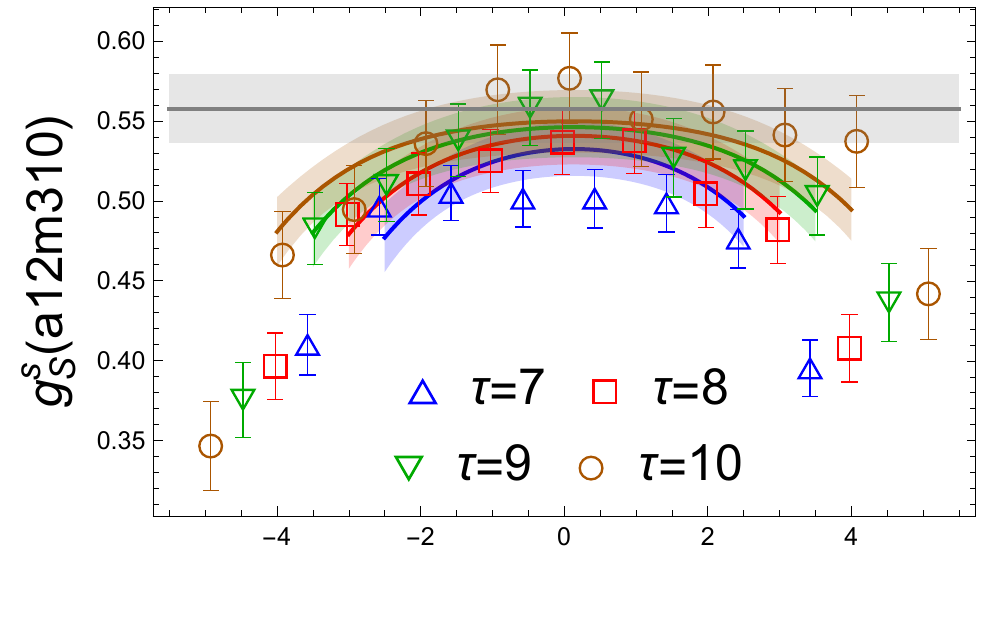}
    \includegraphics[width=0.45\linewidth]{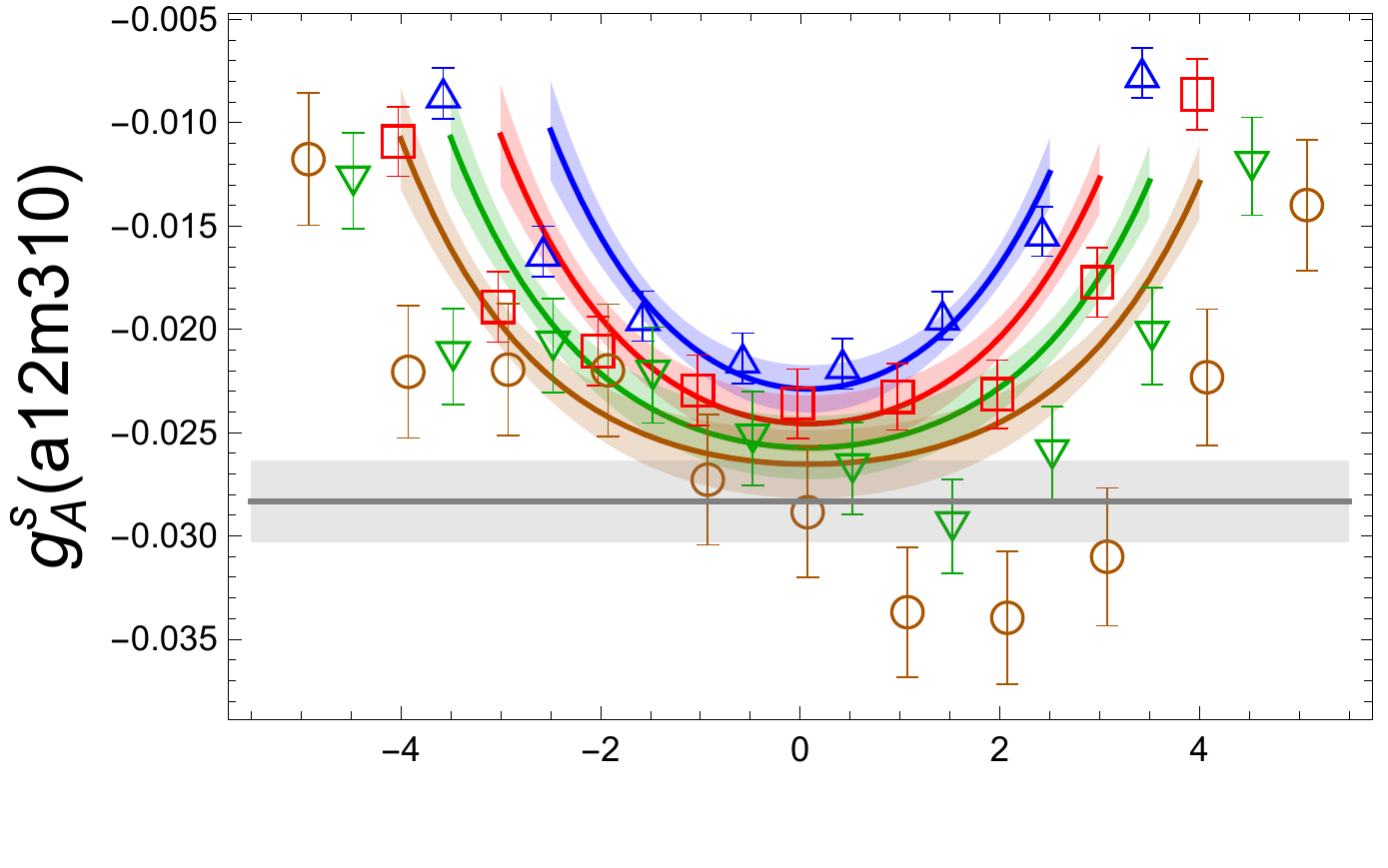}
    \includegraphics[width=0.45\linewidth]{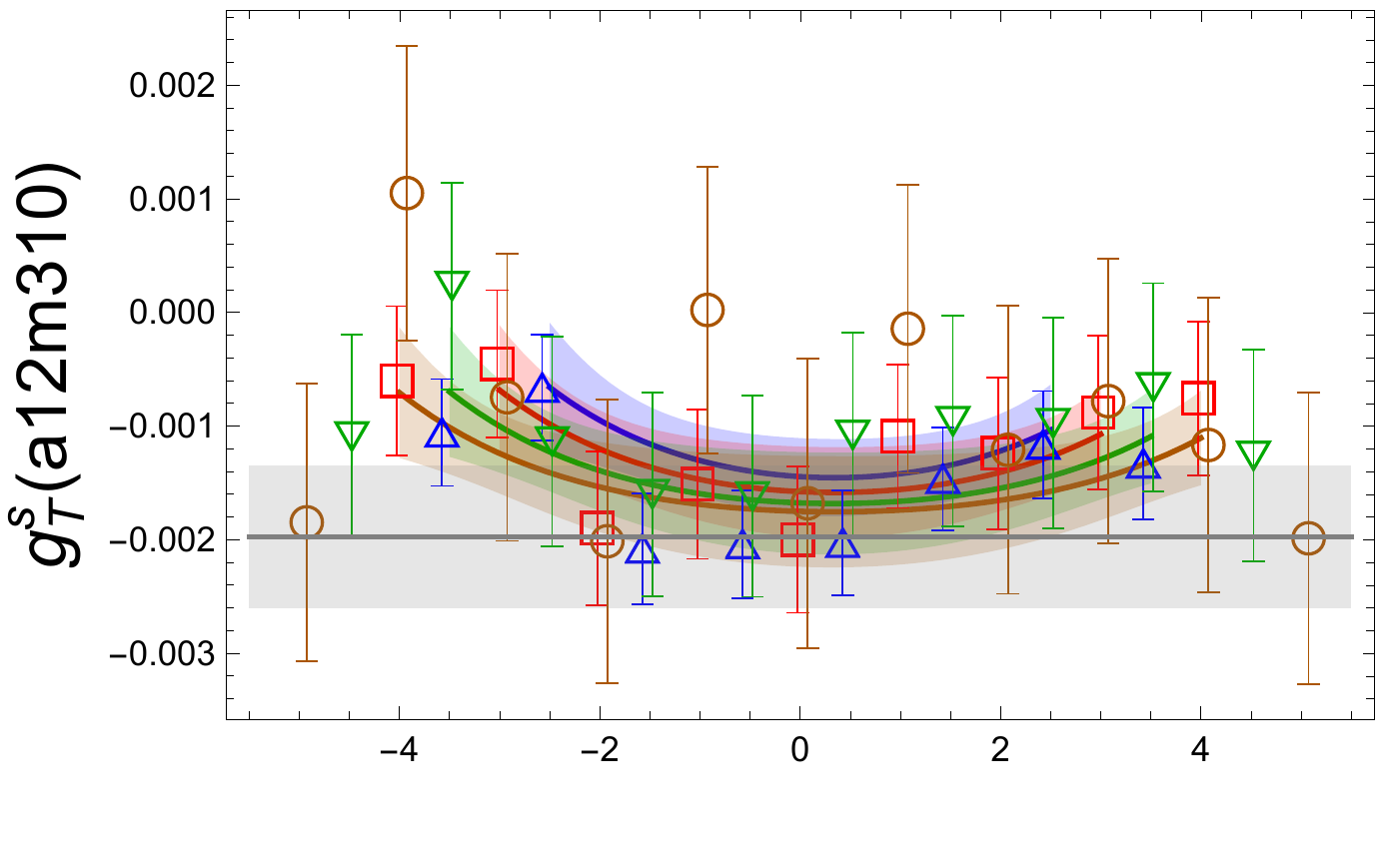}
    \includegraphics[width=0.45\linewidth]{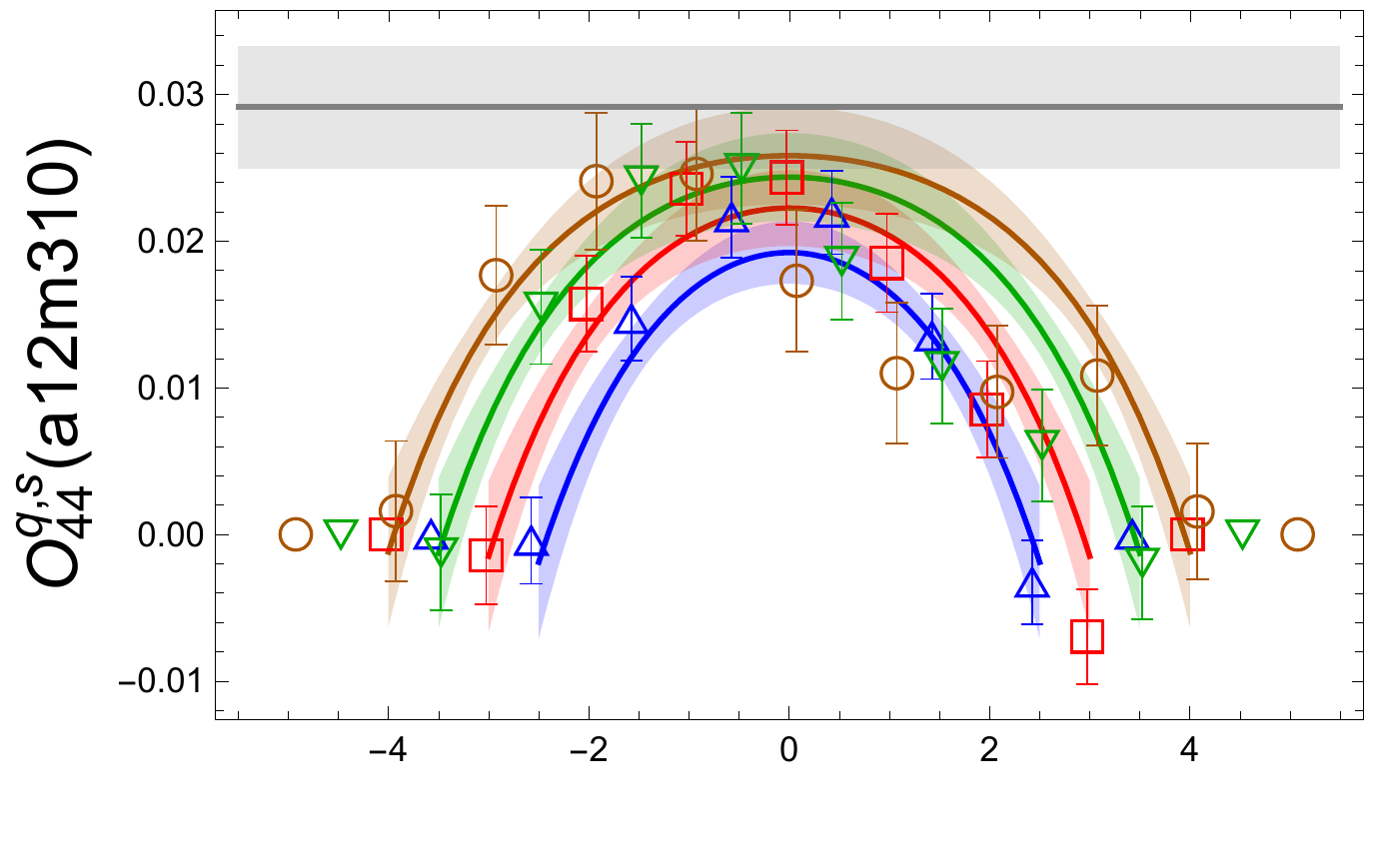}
    \caption{Ratio plots from lattice data (colored markers) and the reconstructed ratio from 2-state fit results (colored bands) on the a12m310 ensemble. The gray bands label the ground state matrix elements.}
    \label{fig:fit}
\end{figure}

We make 2-state fits to the correlators using their spectral decomposition: 
    \begin{equation}
        C^{\mathcal{P}^{unp}}_{\text{2pt}}(\tau)=|A_0|^2 e^{-E_0 \tau}+|A_1|^2e^{-E_1 \tau}
    \end{equation}
    \begin{align}
        C^O_{\text{3pt}}(t,\tau)=\left(|A_0|^2\langle0|O|0\rangle e^{-E_0 t}e^{-E_0 (\tau-t)}+
    A_0A_1\langle1|O|0\rangle e^{-E_0 t}e^{-E_1 (\tau-t)}+\right.\nonumber\\
    \left.A_1A_0\langle0|O|1\rangle e^{-E_1 t}e^{-E_0 (\tau-t)}+
    |A_1|^2\langle1|O|1\rangle e^{-E_1 t}e^{-E_1 (\tau-t)}\right)
    \end{align}
to extract the ground state matrix elements $\langle0|O|0\rangle$. The fit results are displayed  in Fig.~\ref{fig:fit} along with the data for the ratio of 3-point to 2-point functions,  $R(t,\tau)=C^O_{\text{3pt}}(t,\tau)/C^{\mathcal{P}^{unp}}_{\text{2pt}}(\tau)$. The fit results are consistent with the data points.

To renormalize the matrix elements, we used the regularization-independent symmetric momentum subtraction (RI-sMOM) scheme. In this scheme, we choose symmetric momentum configurations for the external quark state and the momentum transfer $p_1^2=p_2^2=(p_2-p_1)^2=\mu_R^2$, as shown in the middle panel of Fig.~\ref{fig:FeymanDiagram}. They are then matched to $\overline{MS}$ scheme using 2-loop perturbation theory, and run to  $\mu=2$~GeV. The $\mu_R$-dependence is then removed by fitting the $Z$-factors to $Z_\Gamma(\mu)=Z_\Gamma+c_1\mu_R^2+c_2\mu_R^4$. The renormalization for the scalar charge and the momentum fraction is more complicated. The momentum fraction for quarks mixes with other flavors as well as gluons, which is not included in our work so far. The scalar charge also mixes with other flavors, as shown in the right of Fig.~\ref{fig:FeymanDiagram}, and we have included the mixing between strange quark and light iso-scalar combination, but not for the charm quark so far.
The two diagrams for calculating $Z_S$ are done for all the flavor combinations as the connected ($ c_{ff'}$) and disconnected ($ d_{ff'}$) parts:
\begin{equation}
    c_{ff'}\equiv\frac{\delta_{ff'}}{12Z^f_\psi}\Tr\left[\langle f|O^{f'}|f\rangle_\text{conn}\right], \,
    d_{ff'}\equiv\frac{-1}{12Z^f_\psi}\Tr\left[\langle f|O^{f'}|f\rangle_\text{disc}\right].
\end{equation} 
Using these, we renormalize the bare charges in the following way:
\begin{equation}
    \begin{pmatrix} g_S^{R,u+d}\\g_S^{R,s}\\g_S^{R,c}\end{pmatrix}=\begin{pmatrix}
c_l-2d_{ll} & -2d_{sl} & -2d_{cl}\\
-d_{ls} & c_s-d_{ss} & -d_{cs}\\
-d_{lc} & -d_{sc} & c_c-d_{cc}
\end{pmatrix}^{-1}\cdot \begin{pmatrix} g_S^\text{bare,u+d}\\g_S^\text{bare,s}\\g_S^\text{bare,c}\end{pmatrix}
\end{equation}
Note that $d_{ff'}$ is sensitive to the flavor in the quark loop, $f'$, but insensitive to the external quark $f$.
\section{Chiral and Continuum Extrapolation}
\begin{figure}
    \centering
    \includegraphics[width=0.45\linewidth]{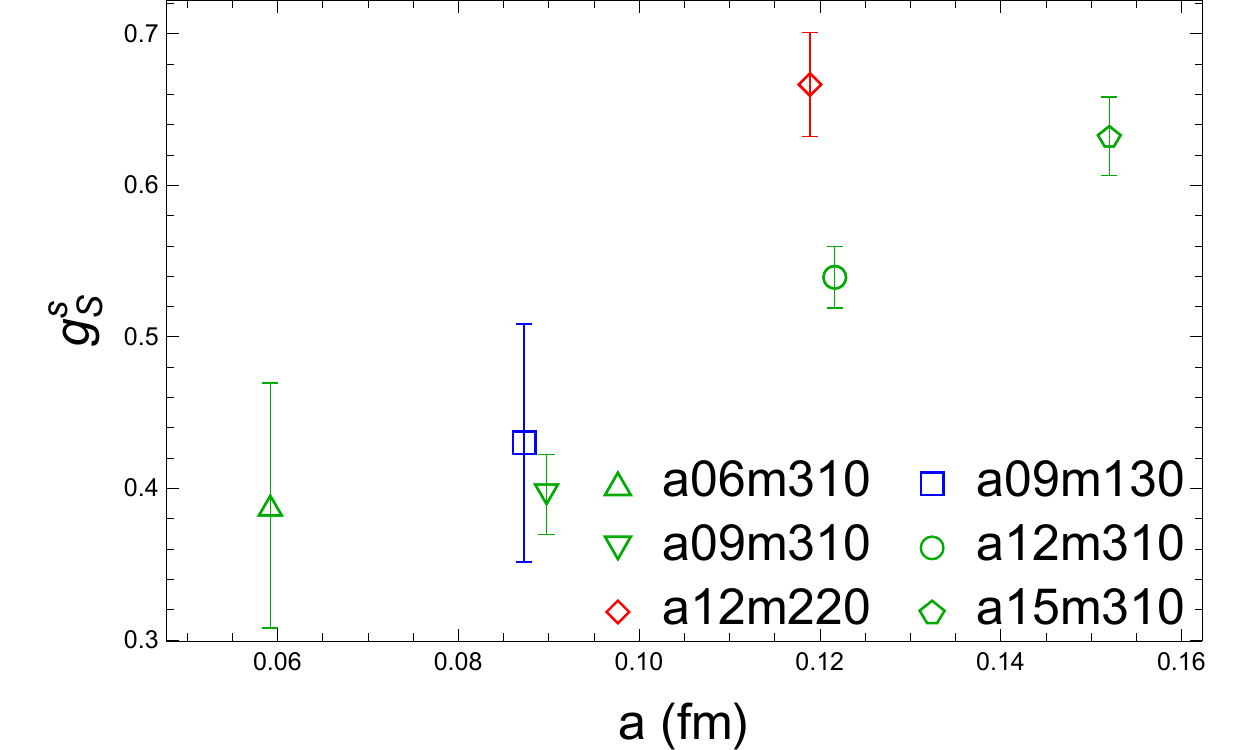}
    \includegraphics[width=0.45\linewidth]{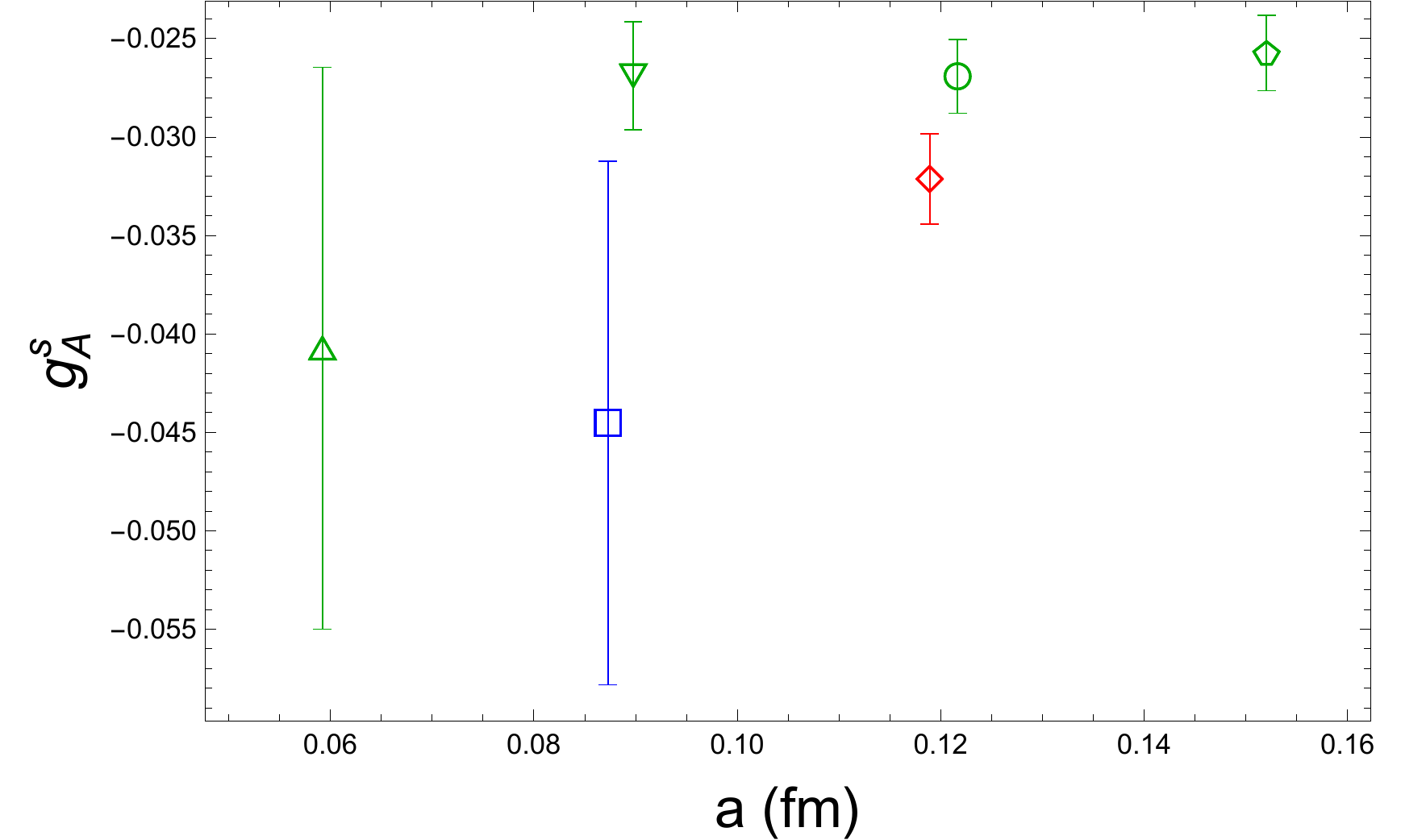}
    \includegraphics[width=0.45\linewidth]{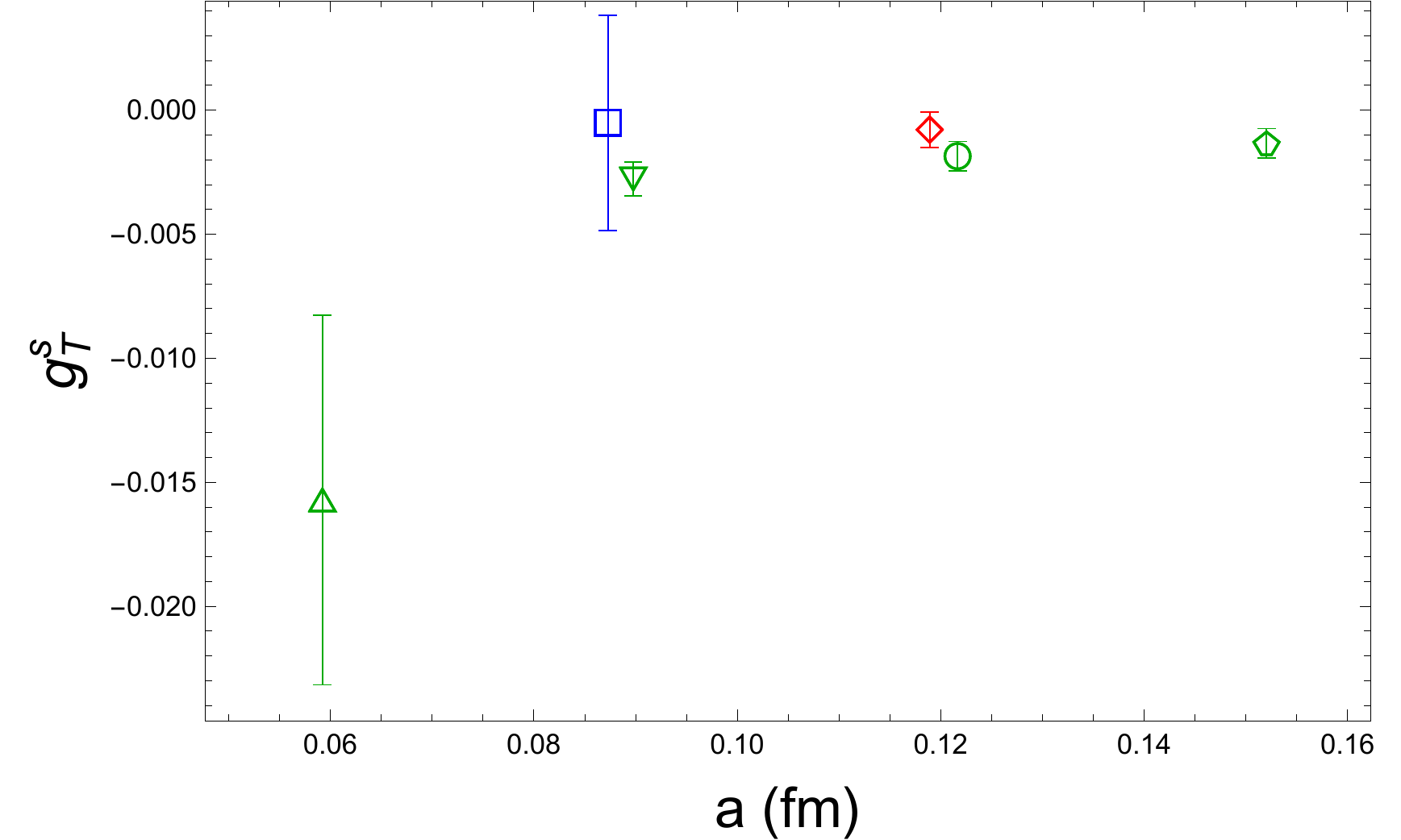}
    \includegraphics[width=0.45\linewidth]{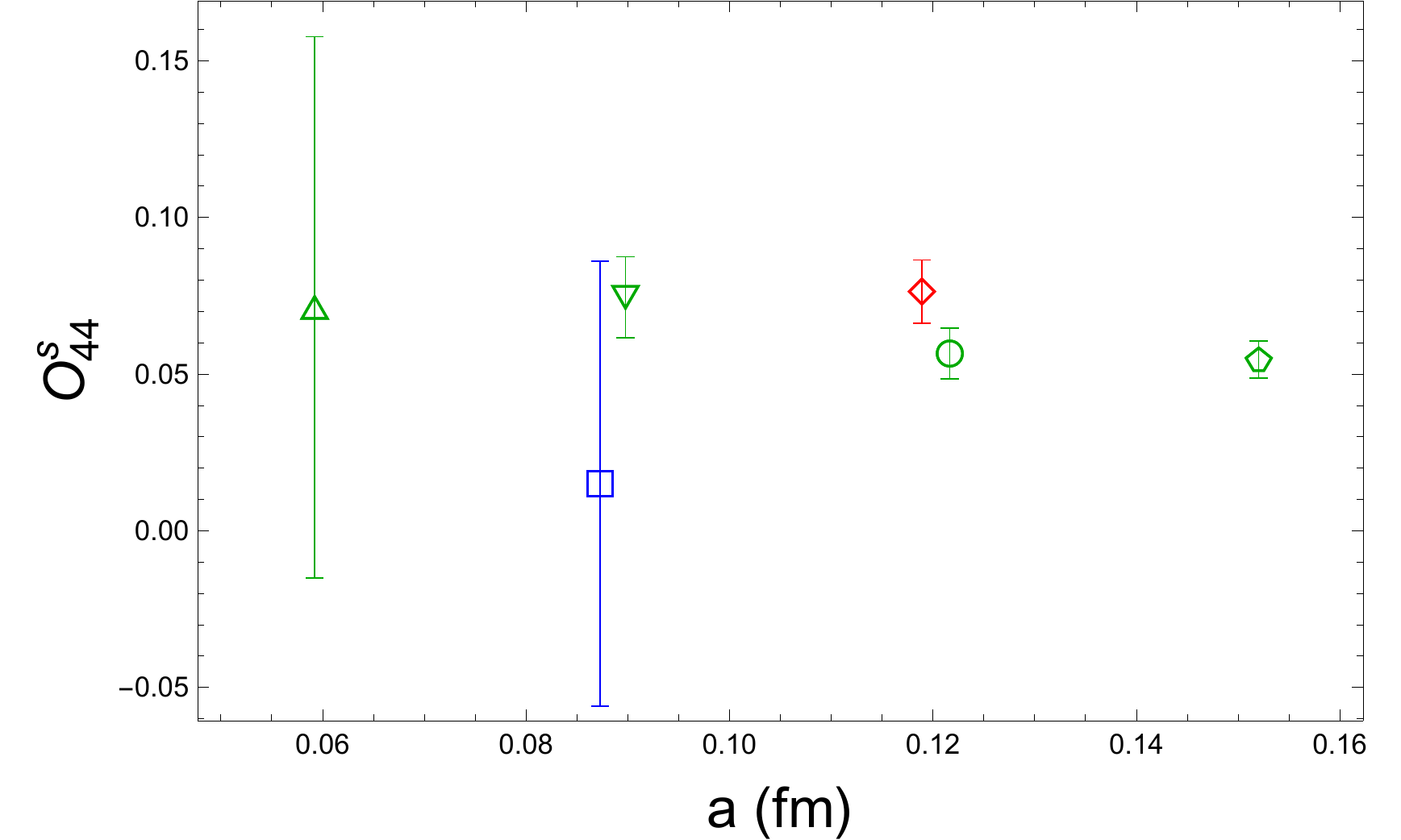}
    \caption{Renormalized charges and momentum fraction for the strange quark.}
    \label{fig:strange_me}
\end{figure}

\begin{figure}
    \centering
    \includegraphics[width=0.45\linewidth]{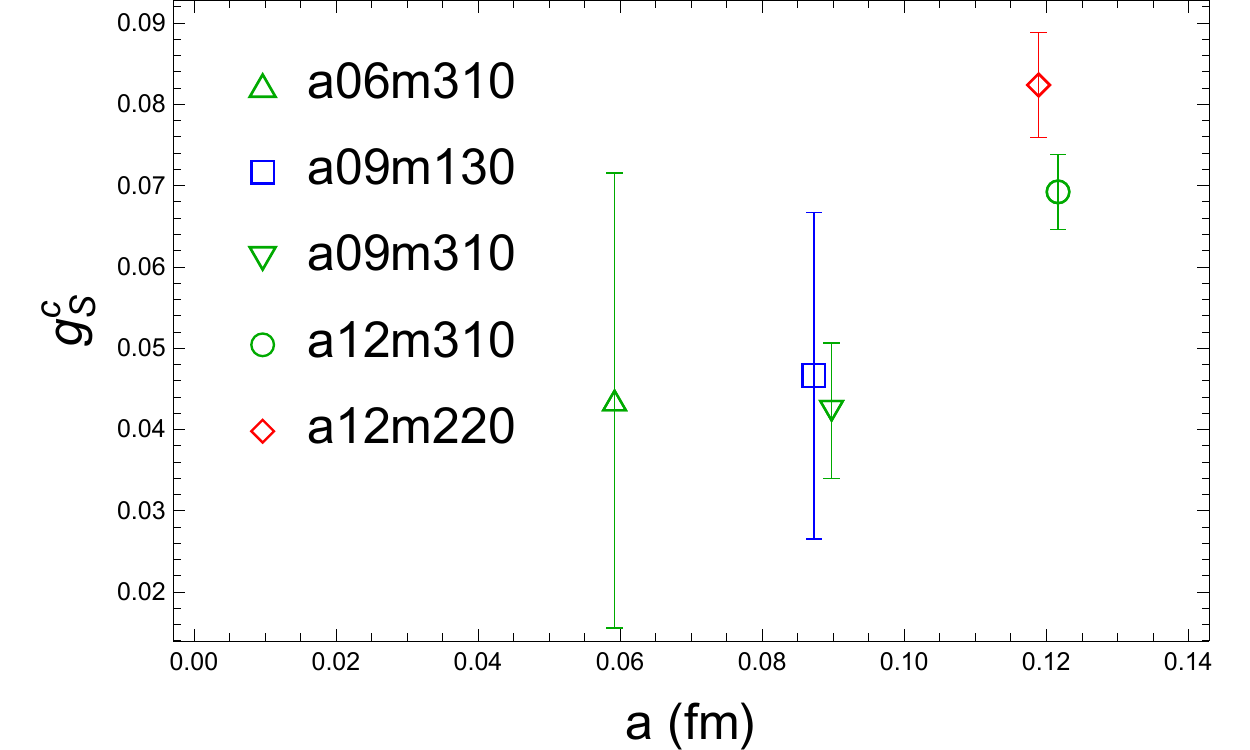}
    \includegraphics[width=0.45\linewidth]{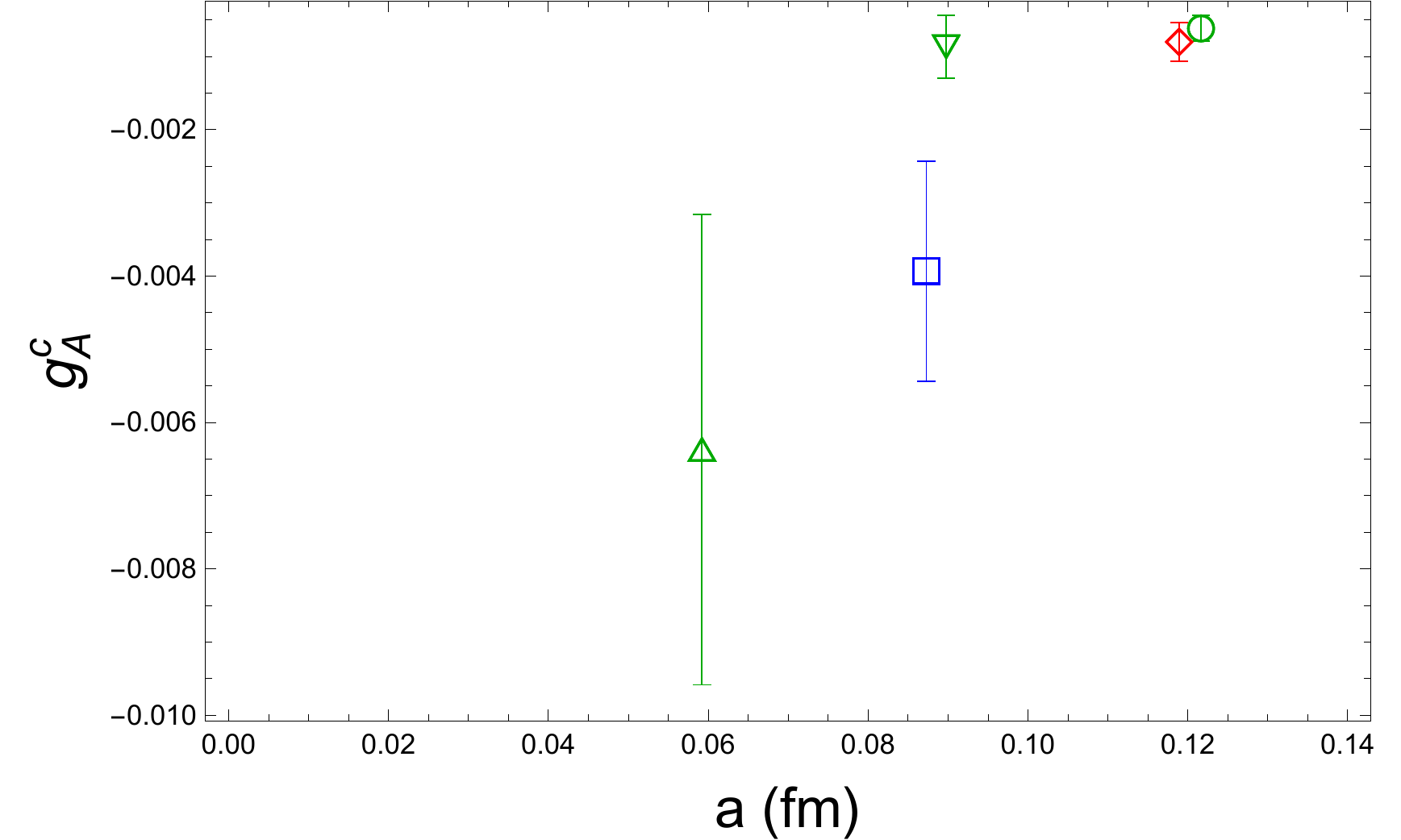}
    \includegraphics[width=0.45\linewidth]{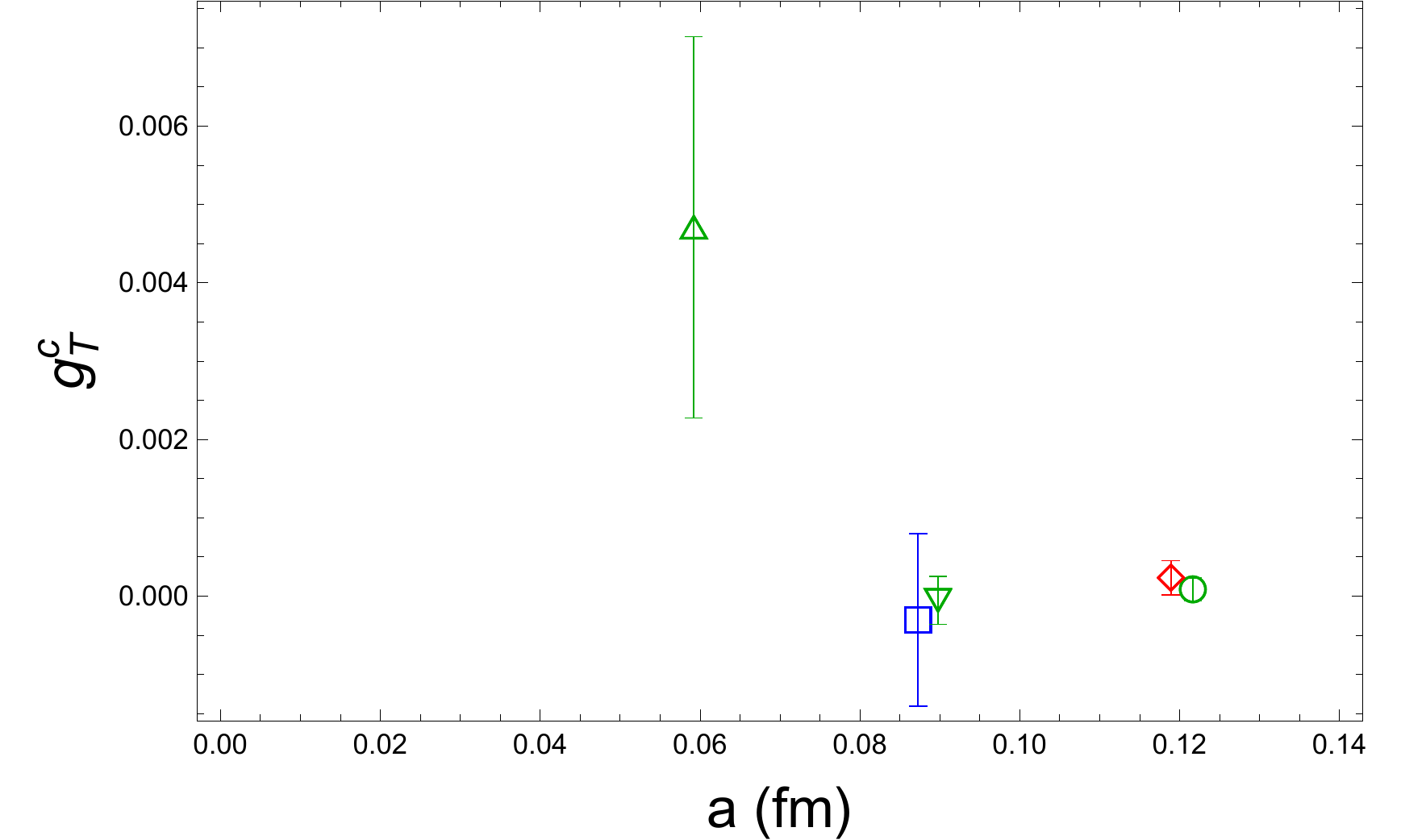}
    \includegraphics[width=0.45\linewidth]{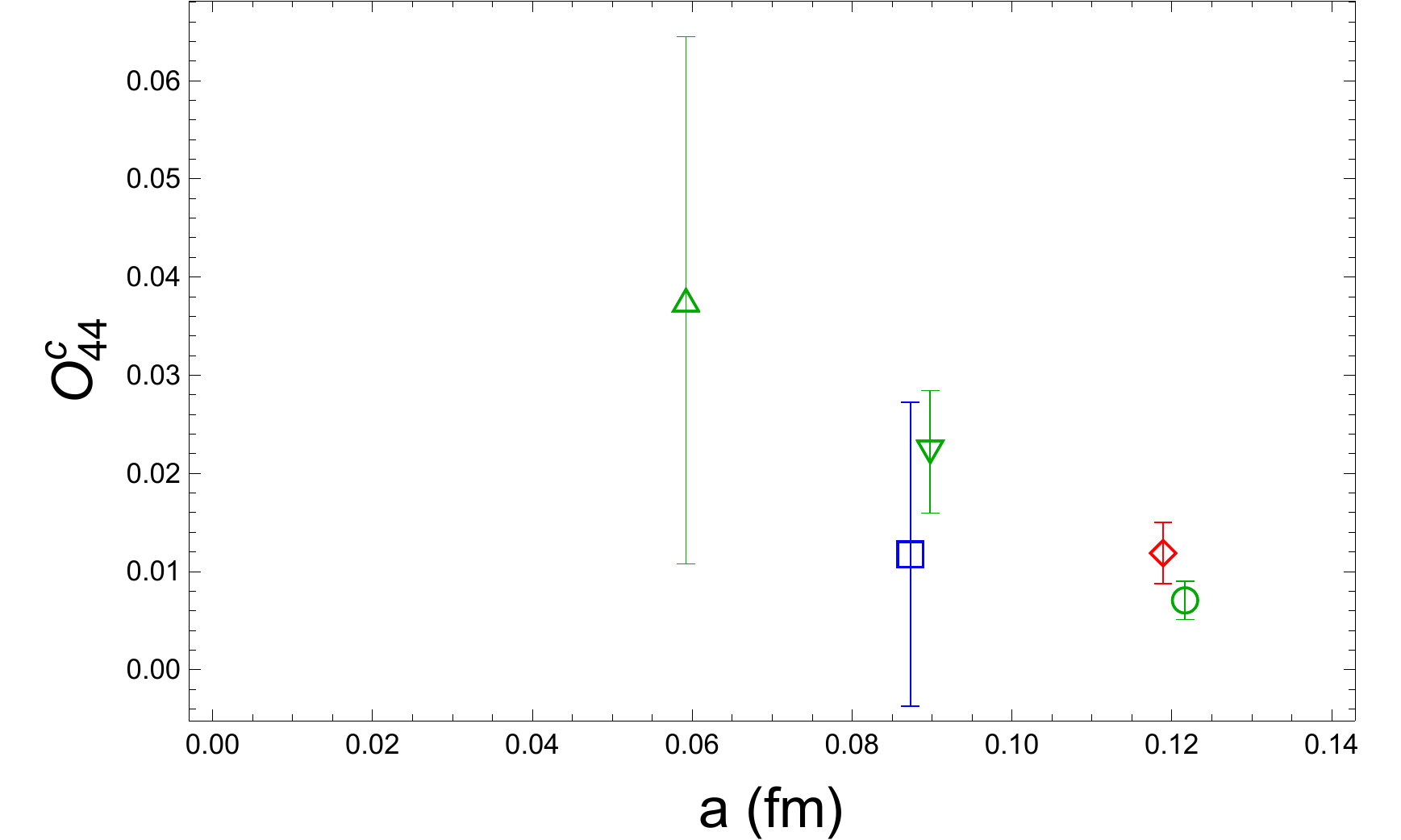}
    \caption{Renormalized charges and momentum fraction for the charm quark.}
    \label{fig:charm_me}
\end{figure}

After renormalization, we obtain the charges and momentum fraction on different ensembles. The results for strange quark are shown in Fig.~\ref{fig:strange_me}, and those for charm quark are shown in Fig.~\ref{fig:charm_me}. We extrapolate these quantities to the physical point by fitting to the ansatz:
\begin{equation}
    g(M_\pi,a)=g^\text{phy}(1+c_ia^i+d_2(M_\pi^2-(M_\pi^\text{phy})^2)),\qquad i=1,2,
    \label{eq:CCfit}
\end{equation}
for linear ($i=1$) or quadratic ($i=2$)  dependence on the lattice spacing.
    
\begin{figure}
    \centering
    \includegraphics[width=0.45\linewidth]{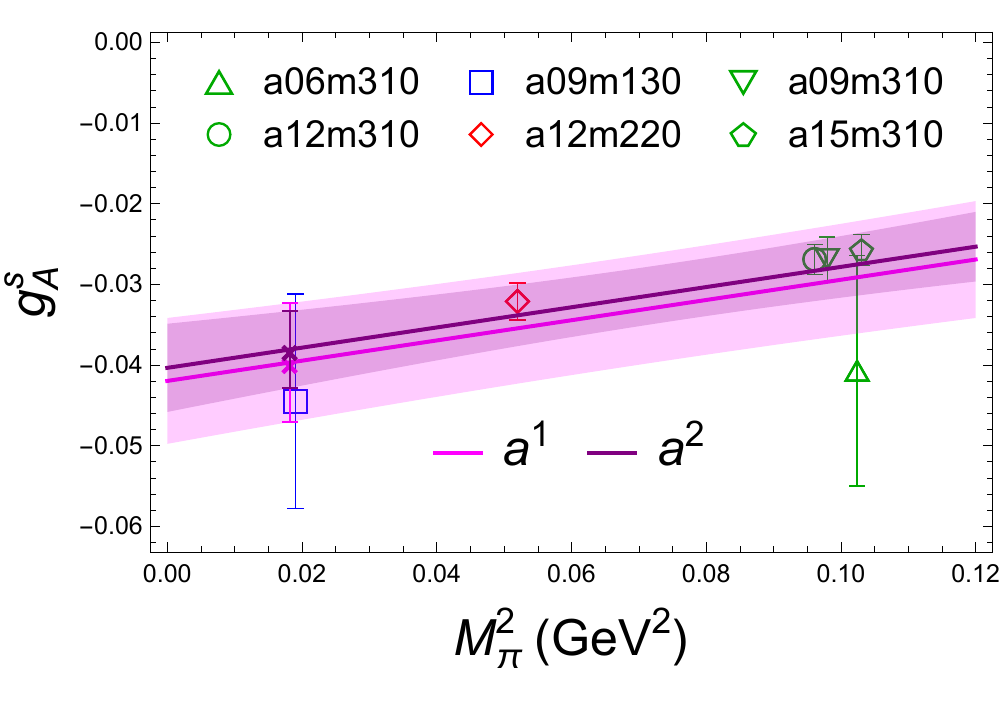}
    \includegraphics[width=0.45\linewidth]{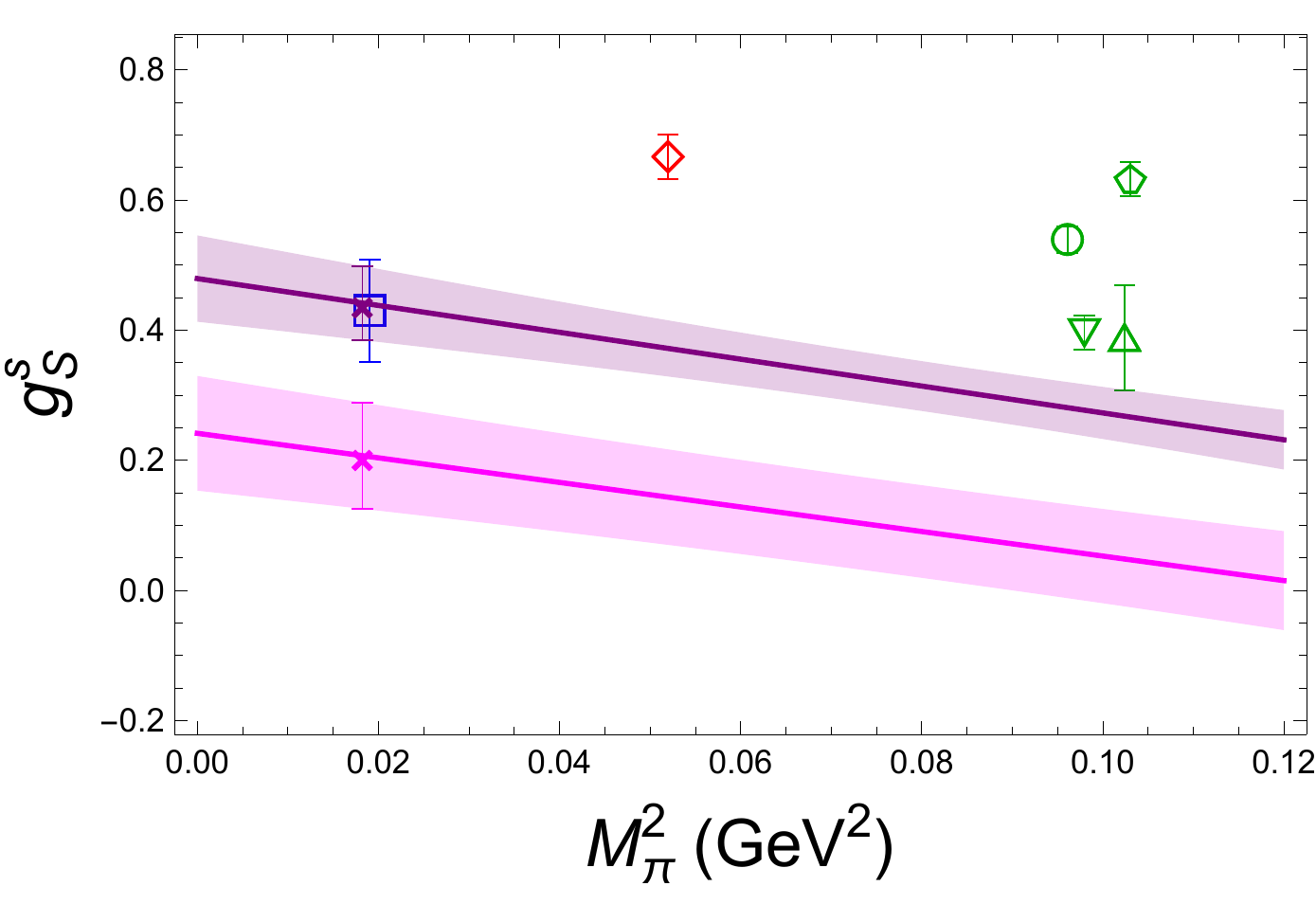}
    \includegraphics[width=0.45\linewidth]{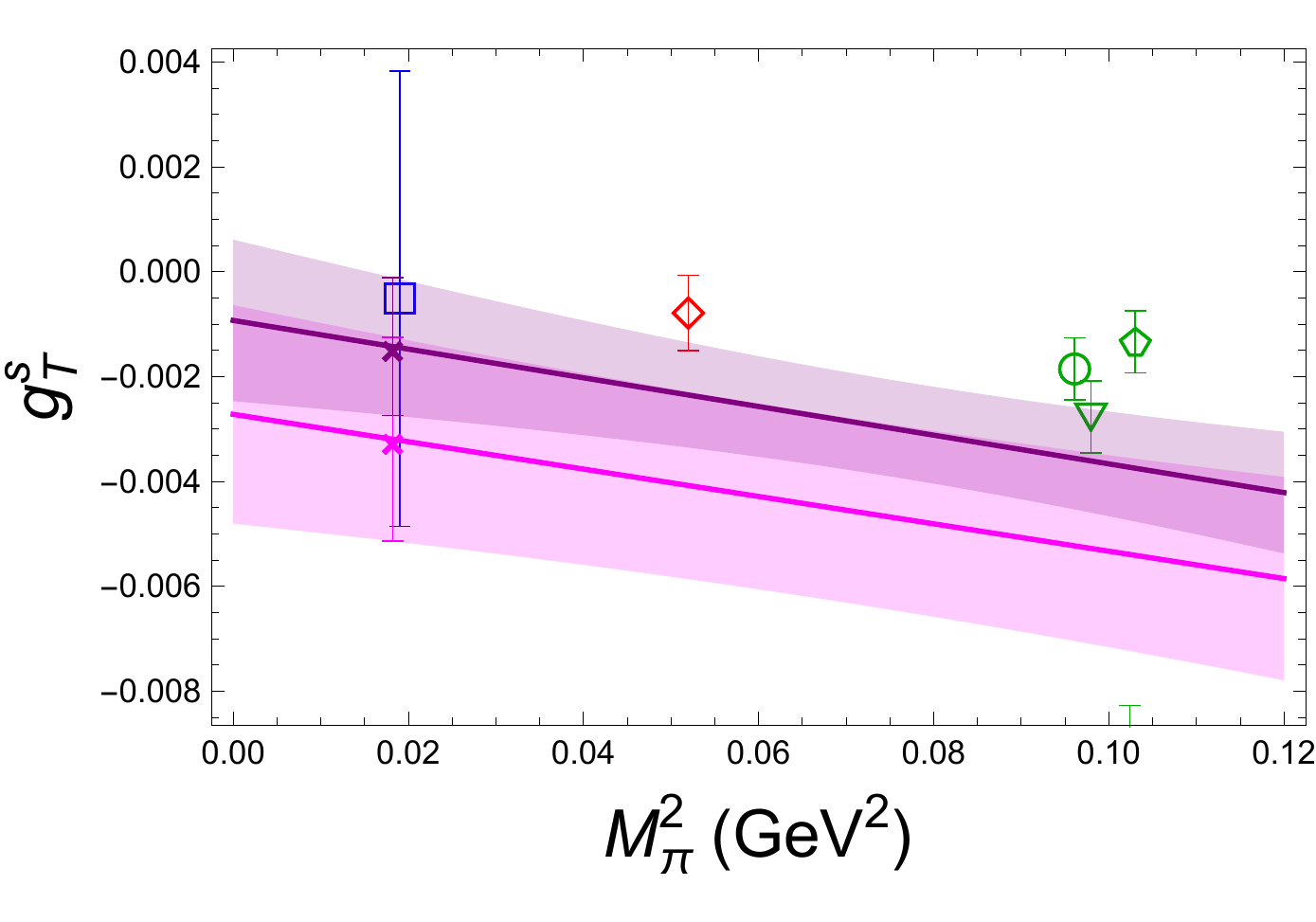}
    \includegraphics[width=0.45\linewidth]{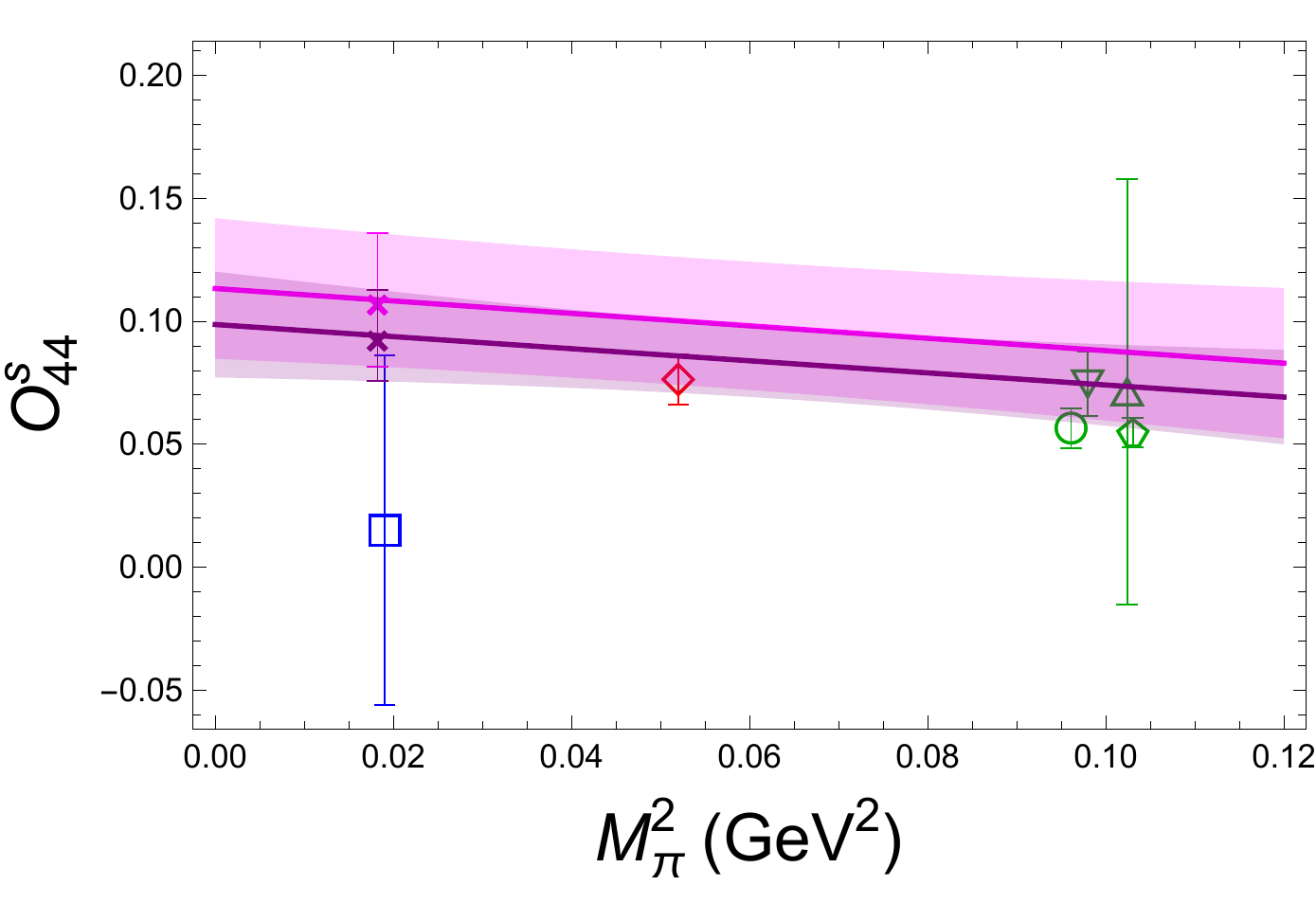}
    \caption{Chiral and continuum extrapolation of strange contribution to charges and momentum fraction using Eq.~\protect\eqref{eq:CCfit}, and shown as a function of $M_\pi^2$ with $a=0$.}
    \label{fig:strange_extrapolate_mass}
\end{figure}

\begin{figure}
    \centering
    \includegraphics[width=0.45\linewidth]{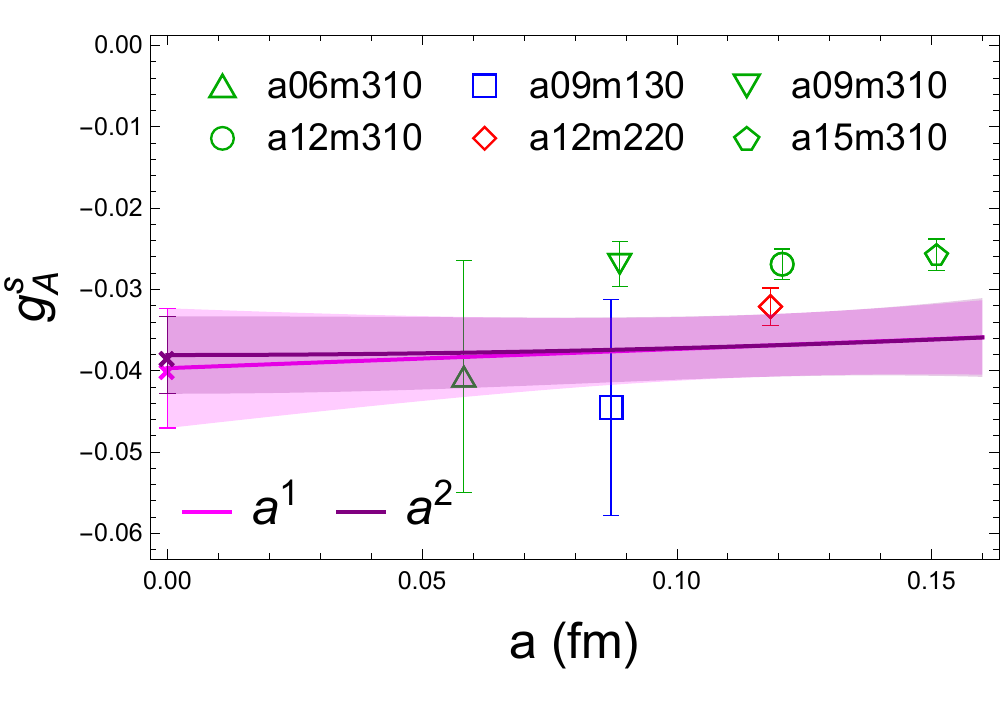}
    \includegraphics[width=0.45\linewidth]{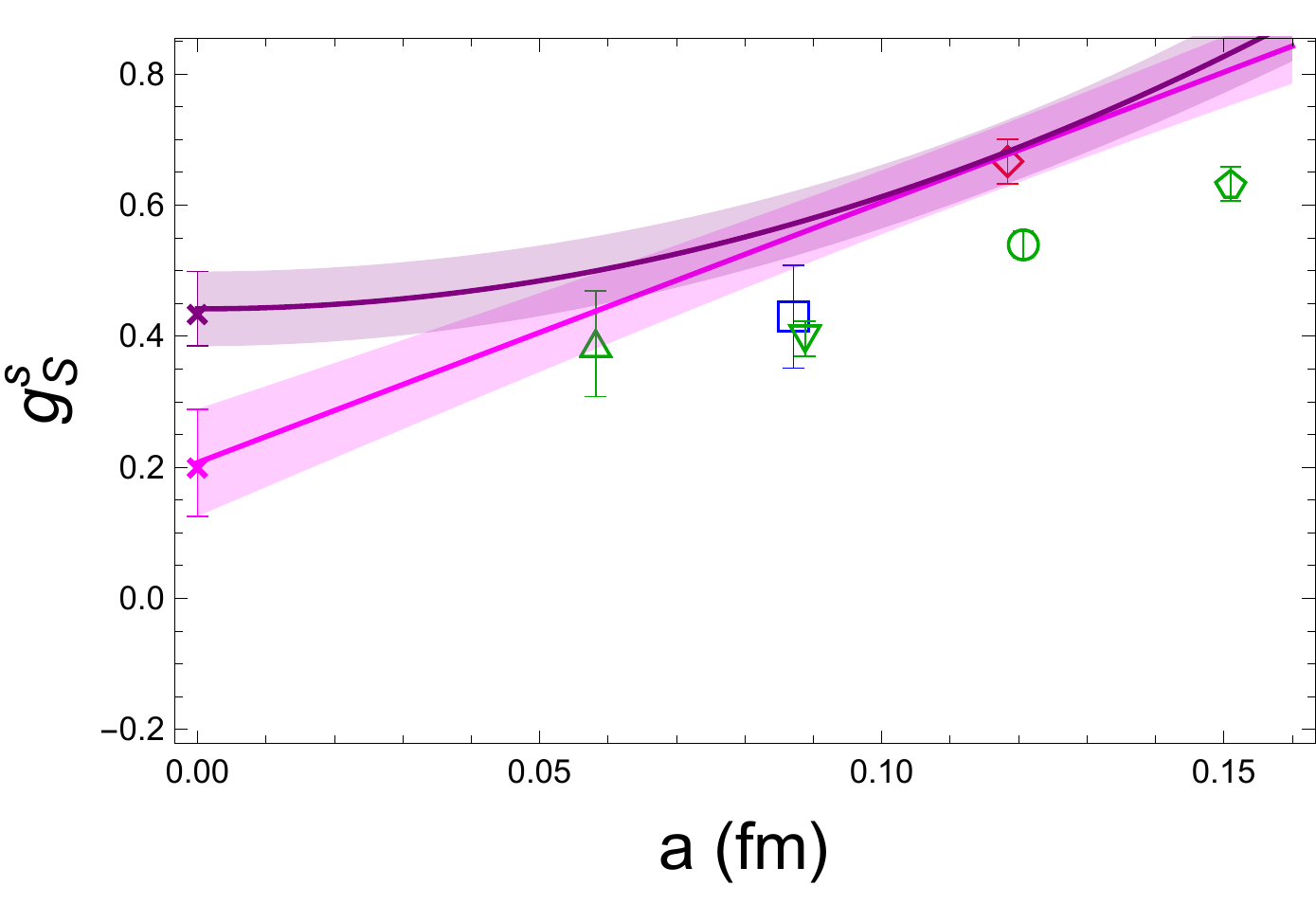}
    \includegraphics[width=0.45\linewidth]{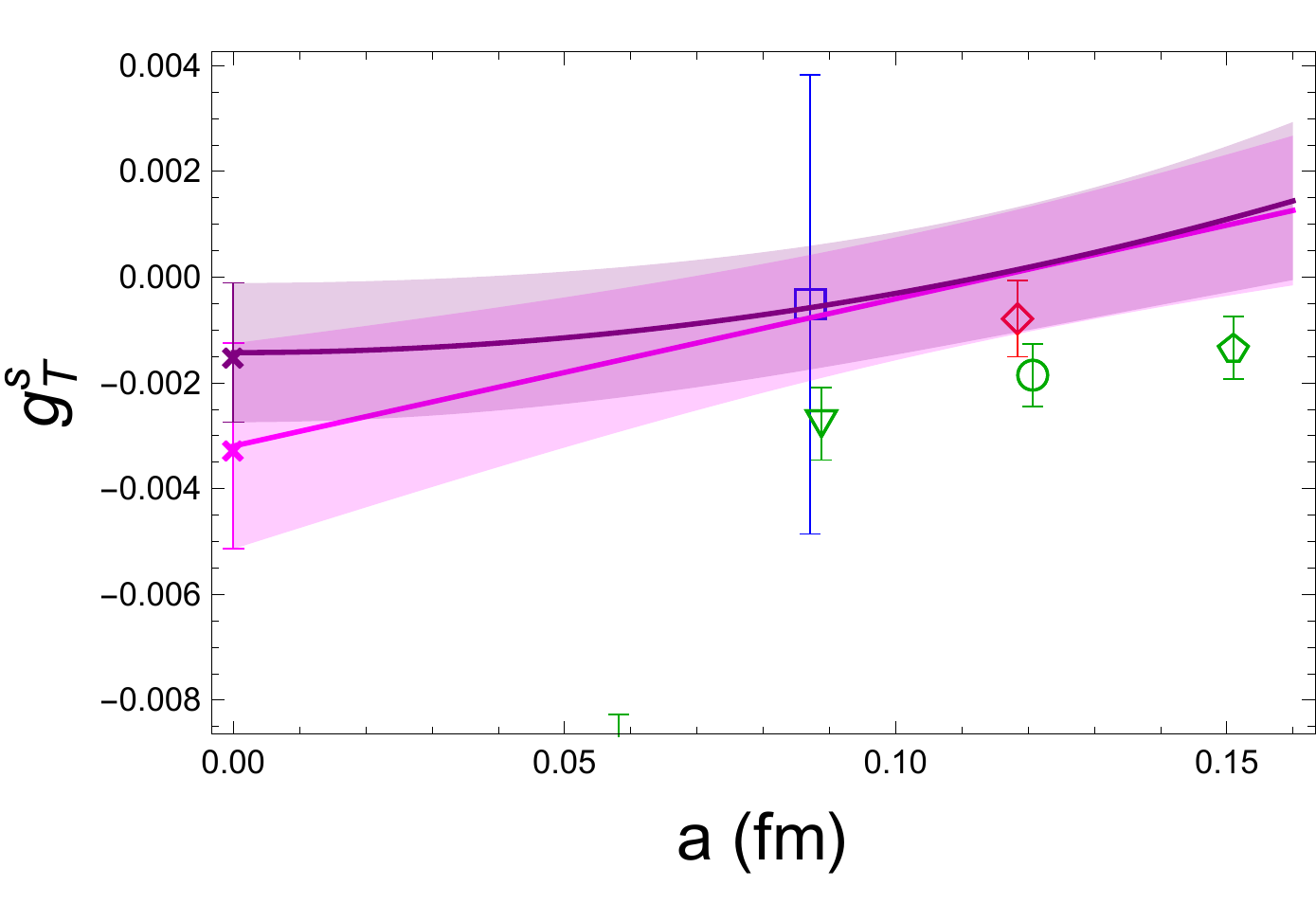}
    \includegraphics[width=0.45\linewidth]{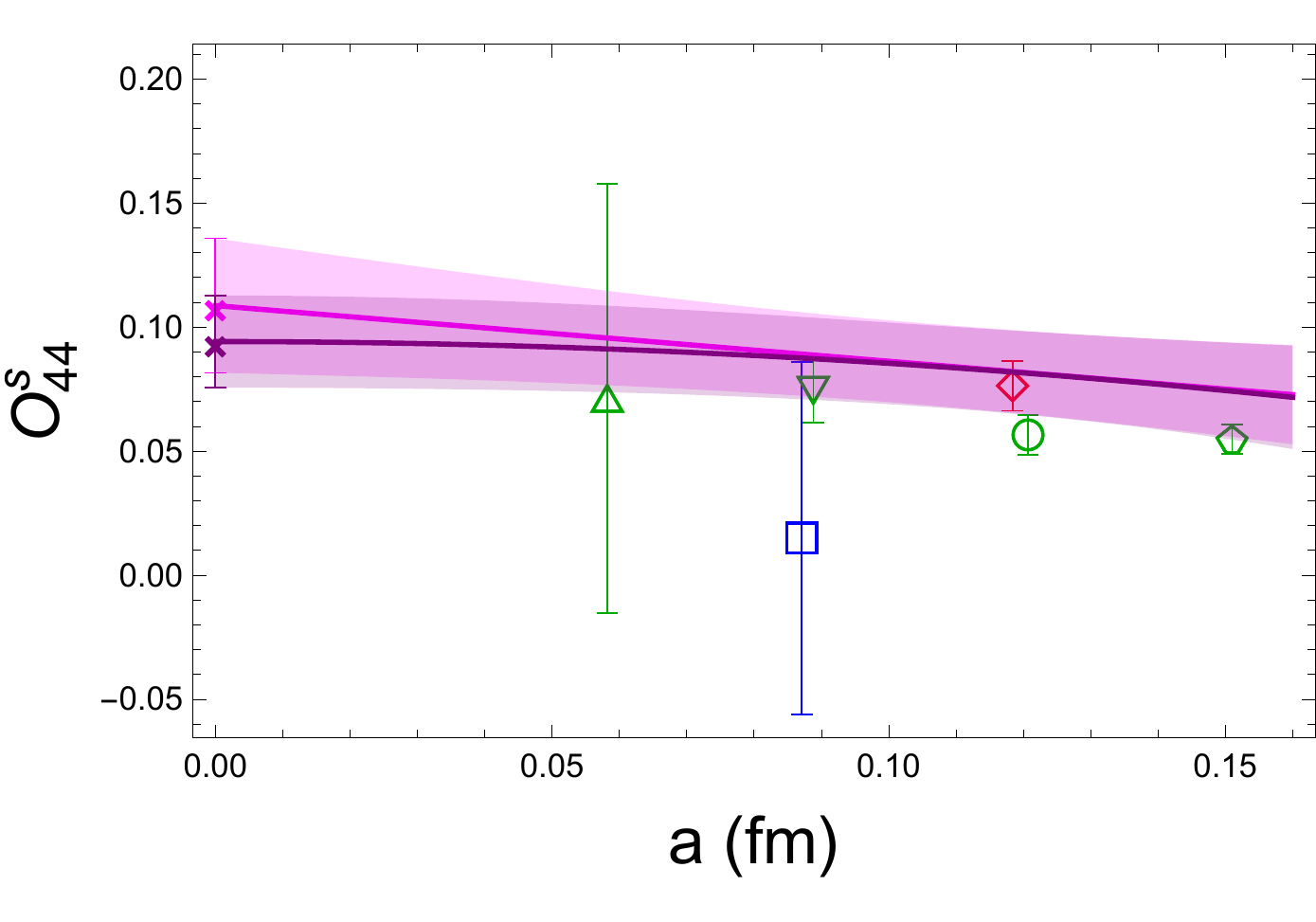}
\caption{Chiral and continuum extrapolation of strange contribution to charges and momentum fraction using Eq.~\protect\eqref{eq:CCfit}, and shown as a function of lattice spacing with $M_\pi$ set to $135$~MeV. }
\label{fig:strange_extrapolate_spacing}
\end{figure}

The extrapolations linear (quadratic) in lattice spacing are shown as magenta (purple) bands in Fig.~\ref{fig:strange_extrapolate_mass} and Fig.~\ref{fig:strange_extrapolate_spacing} for strange quark results. We notice that the $M_\pi$ dependence is non-negligible while the $a$ dependence in $g_A$, $g_T$, $O_{44}^S$ is not clear. The $a$ dependence is large in $g_S$. The linear and quadratic fits have similar $\chi^2$ and the two results are consistent within 1$\sigma$ except for $g_S$. We are working on improving the statistics on finer lattices to better constrain the extrapolation.


\begin{figure}
    \centering
    \includegraphics[width=0.45\linewidth]{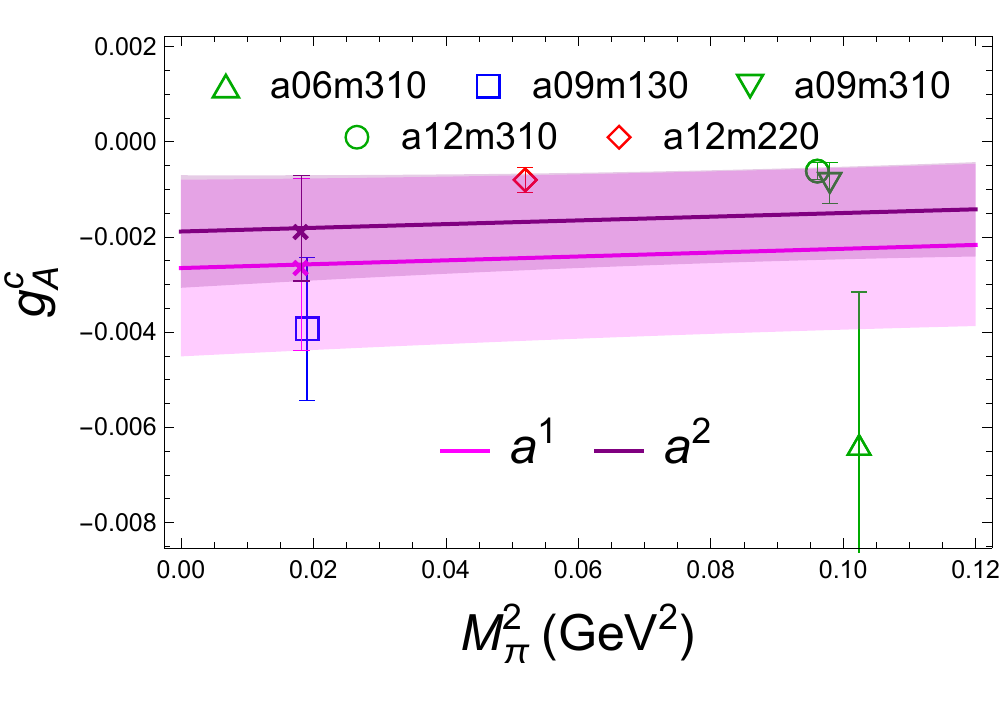}
    \includegraphics[width=0.45\linewidth]{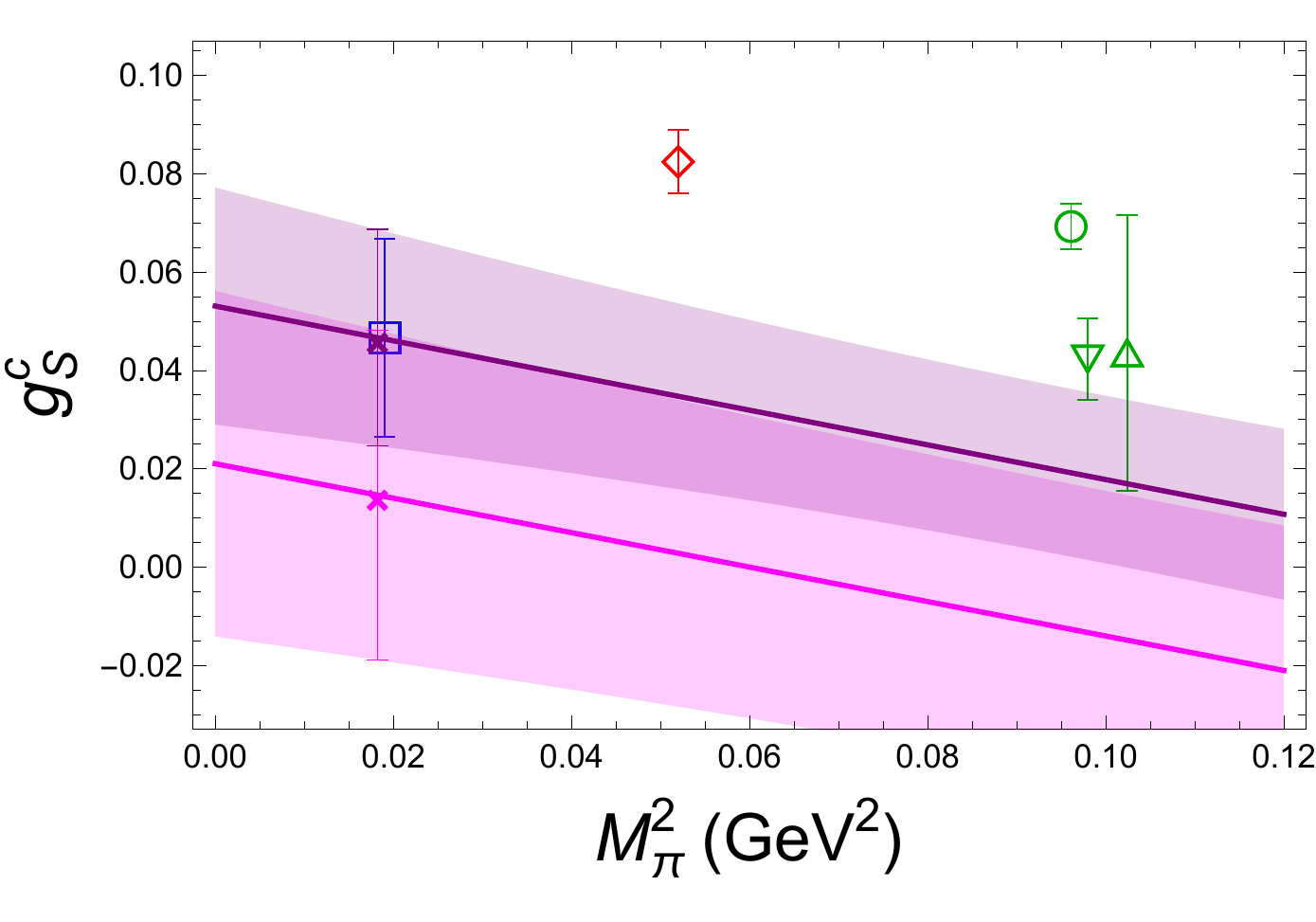}
    \includegraphics[width=0.45\linewidth]{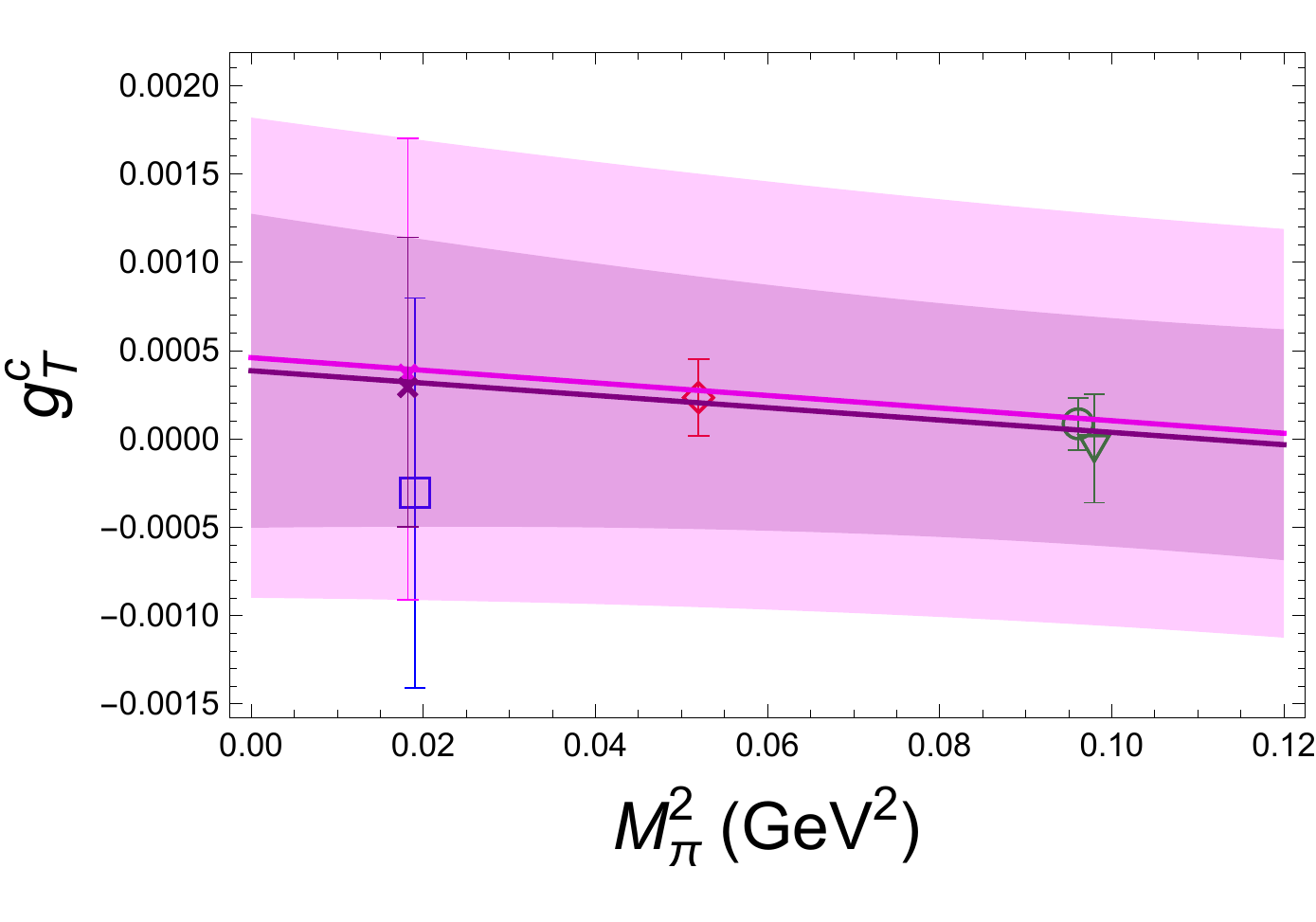}
    \includegraphics[width=0.45\linewidth]{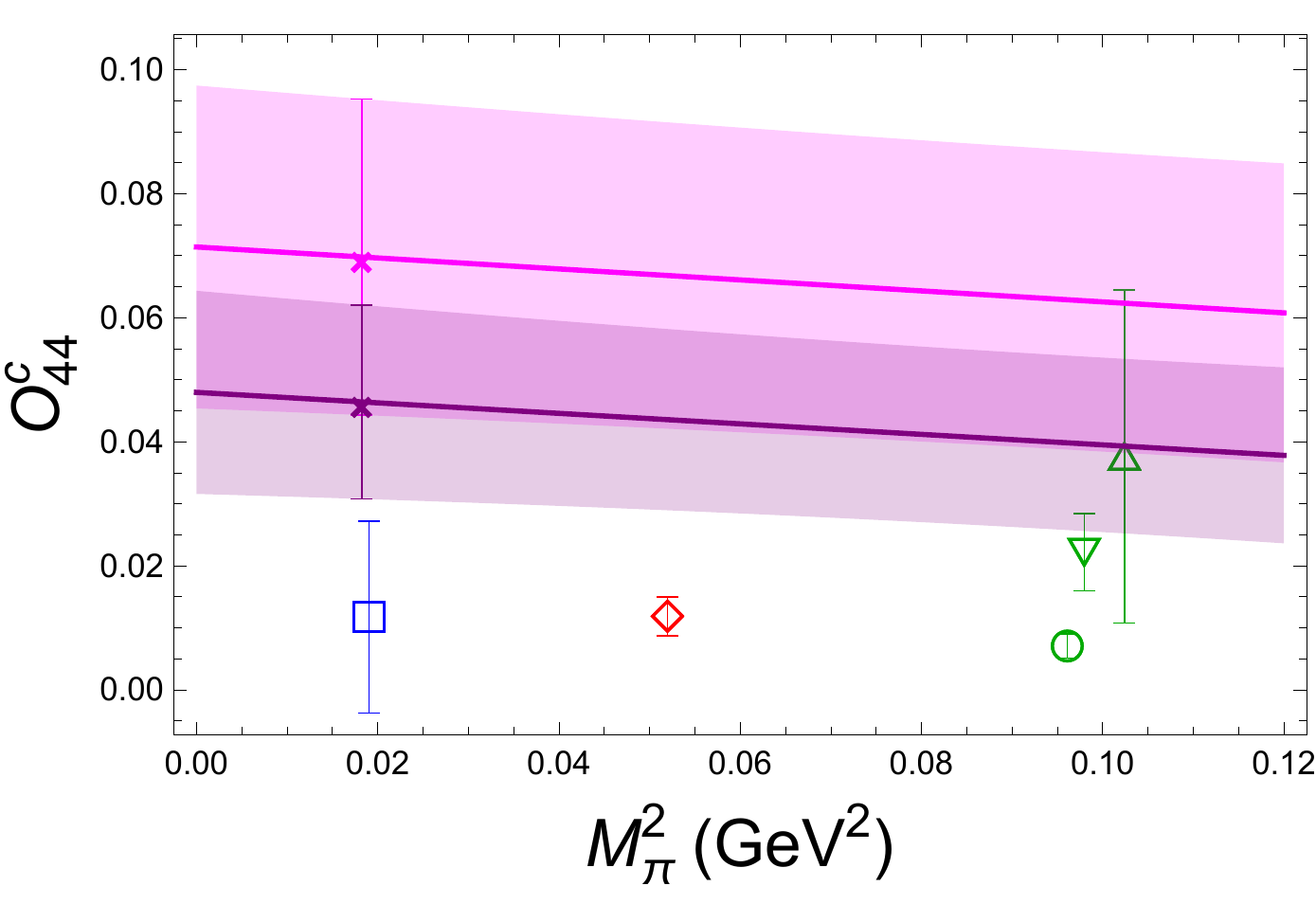}
    \caption{Chiral and continuum extrapolation of charm contribution to charges and momentum fraction using Eq.~\protect\eqref{eq:CCfit}, and shown as a function of $M_\pi^2$ with $a=0$.}
    \label{fig:charm_extrapolate_mass}
\end{figure}

\begin{figure}
    \centering
    \includegraphics[width=0.45\linewidth]{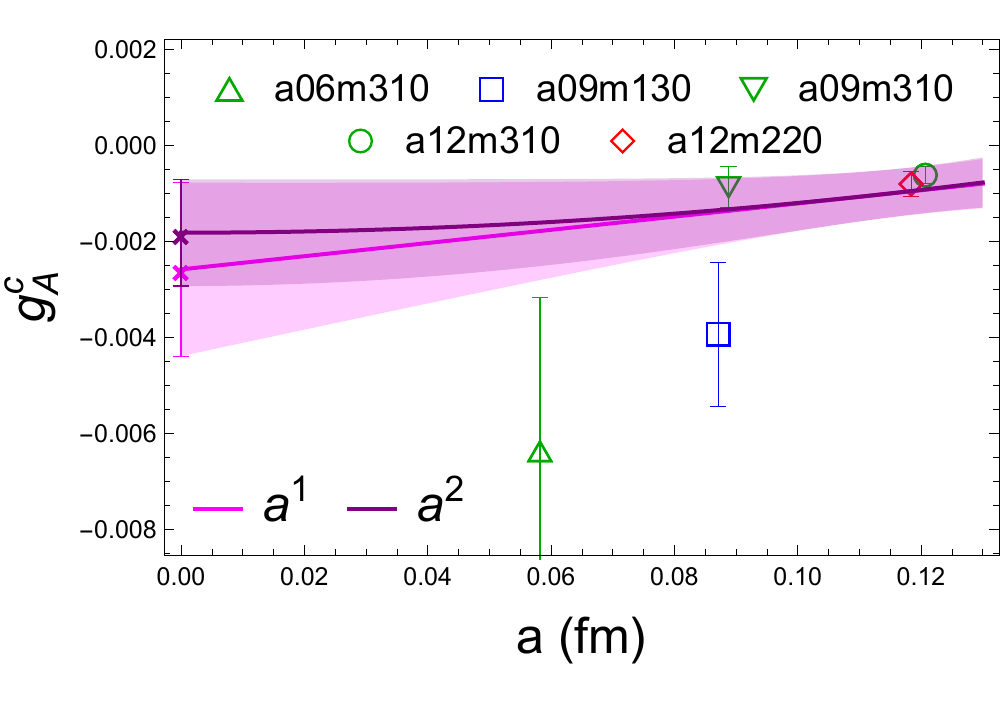}
    \includegraphics[width=0.45\linewidth]{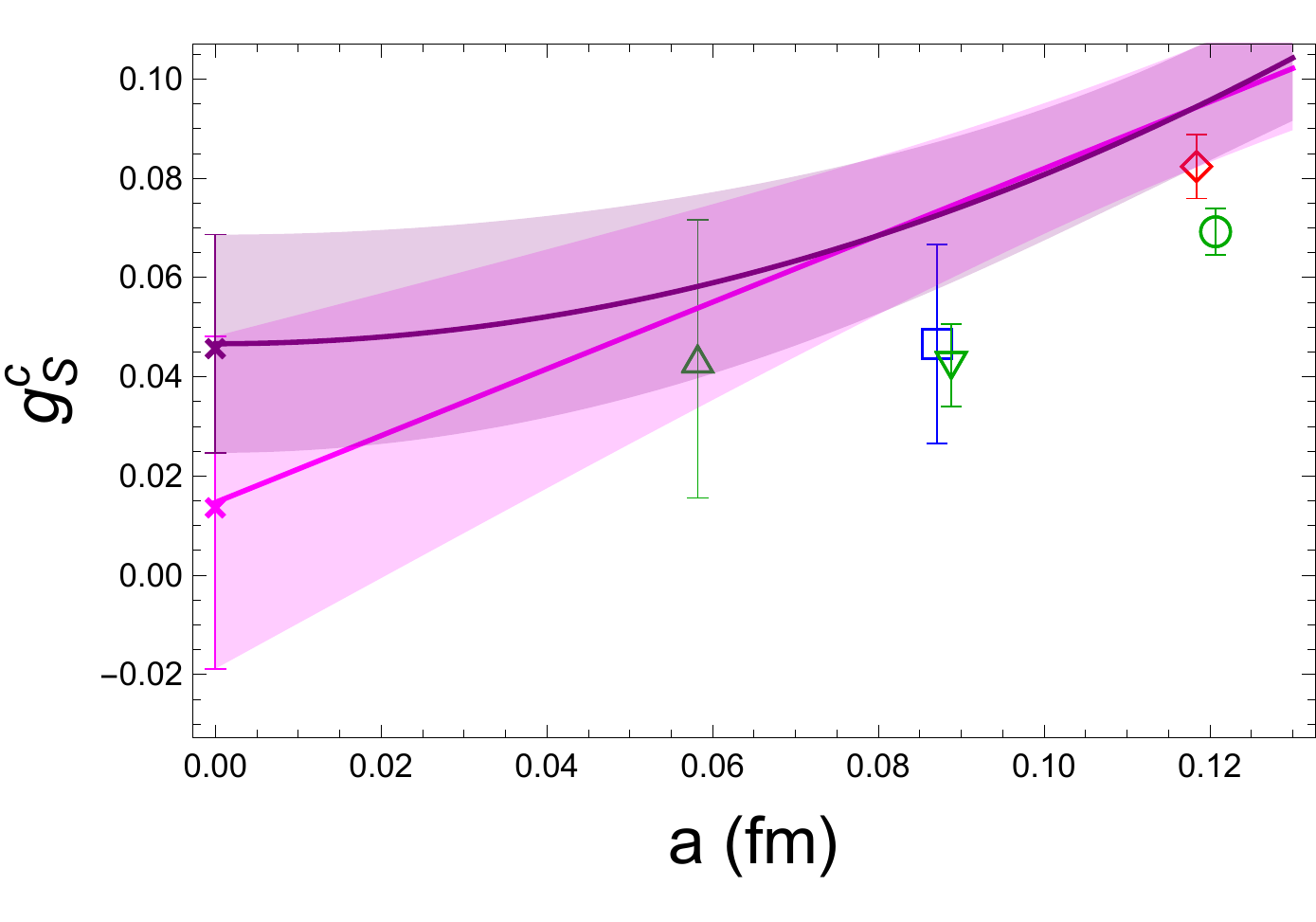}
    \includegraphics[width=0.45\linewidth]{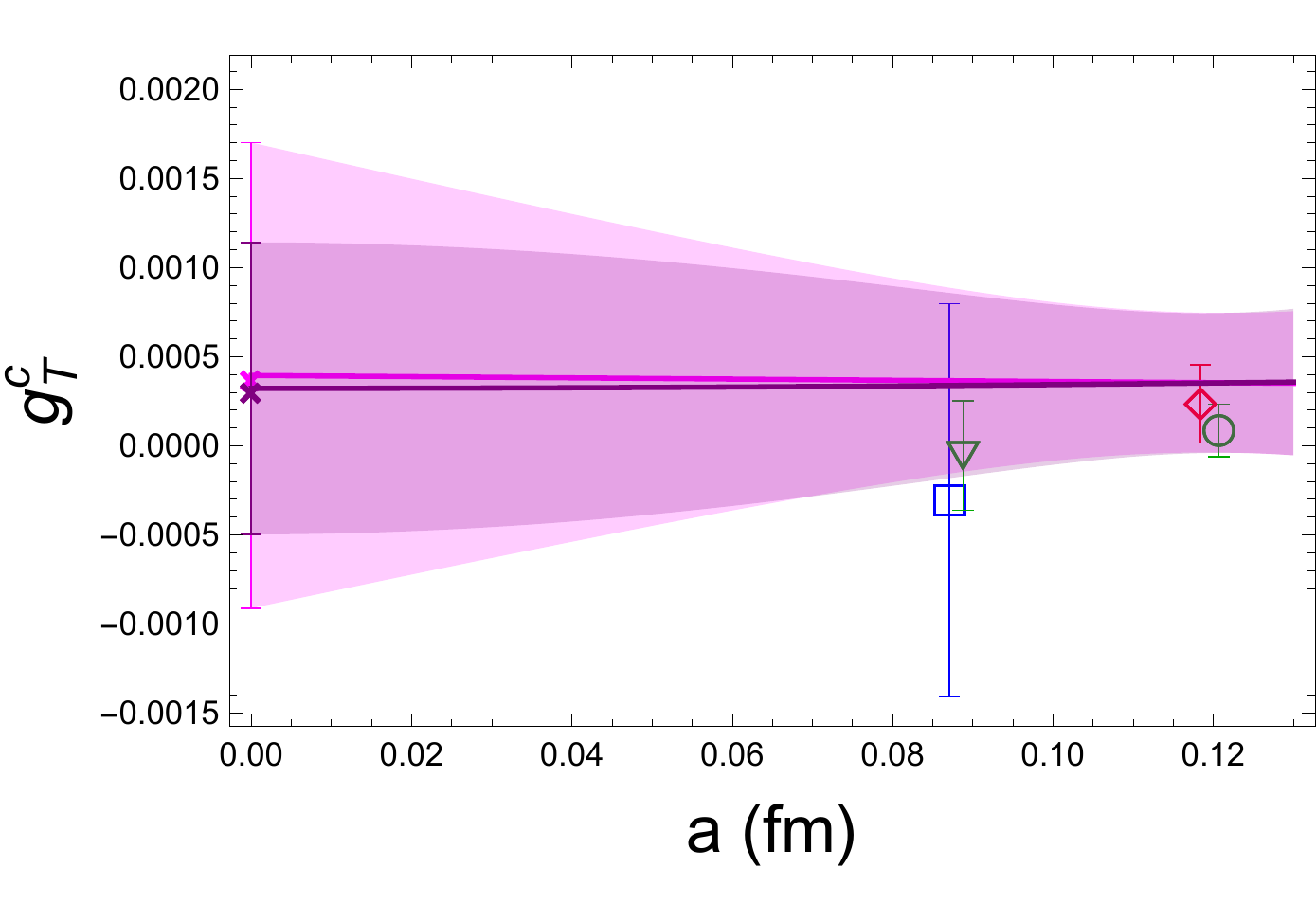}
    \includegraphics[width=0.45\linewidth]{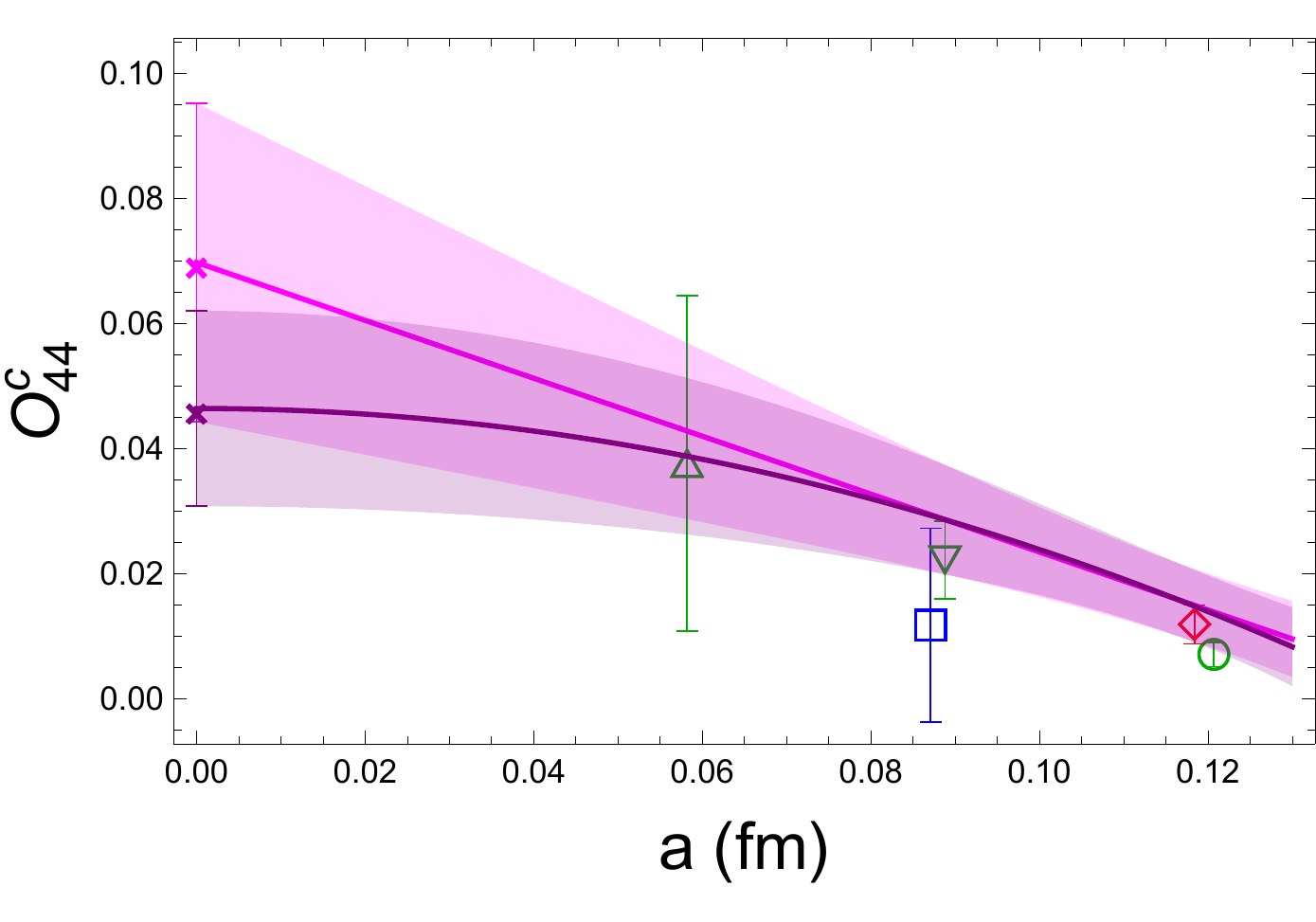}
\caption{Chiral and continuum extrapolation of charm contribution to charges and momentum fraction using Eq.~\protect\eqref{eq:CCfit}, and shown as a function of lattice spacing, with $M_\pi$ set to $135$~MeV.}
\label{fig:charm_extrapolate_spacing}
\end{figure}

The extrapolations linear (quadratic) in lattice spacing are shown as magenta (purple) bands in Fig.~\ref{fig:charm_extrapolate_mass} and Fig.~\ref{fig:charm_extrapolate_spacing} for charm quark results. Unlike the strange results, the $M_\pi$ dependence is not significant for charm. The $a$ and $a^2$ extrapolations are consistent within 1$\sigma$. We also note that the  scalar charge reduces under continuum extrapolation, while the momentum fraction becomes larger under continuum extrapolation.


Comparison is made with previous results in Tab.~\ref{tab:compare_strange} for the strange quark and in Tab.~\ref{tab:compare_charm} for the charm quark. We find that the axial and tensor charges are consistent with previous results. Our scalar charge has large uncertainty because of the smaller statistics on finer lattices. The momentum fraction (without including mixing correction) is much larger.
\begin{table}
    \centering
    \begin{tabular}{|c|c|c|c|c|}
    \hline
          strange &  $g^{s,R}_S$ & $g^{s,R}_A$ &$g^{s,R}_T$ & $\langle x\rangle_s$ \\
         \hline
         \!This work (preliminary)\!&\!\!$0.13\sim0.50$\!\!\!& $-0.0397(73)$ & $-0.0032(19)$ & $0.109(27)$ \\
         \hline
         PNDME'18~\cite{Gupta:2018lvp,Lin:2018obj} & n/a &$-0.053(8)$&$-0.0027(16)$ & n/a\\
         \hline
         $\star$ETMC'19~\cite{Alexandrou:2019brg,Alexandrou:2020sml} Nf=4&$0.454(16)$ &$-0.0458(73)$ &$-0.00268(58)$ & 0.052(12)\\
         \hline
         $\star$ETMC'19~\cite{Alexandrou:2019brg} Nf=2&$0.371(38)$ &$-0.061(17)$ &$-0.0041(12)$ & n/a\\
         \hline
         $\chi$QCD'18~\cite{Yang:2018nqn,Liang:2018pis} & n/a &$-0.035(9)$ & n/a & 0.051(26)\\
         \hline
         $\star$JLQCD'18~\cite{Yamanaka:2018uud} &$0.15(16)$ &$-0.046(28)$ & $-0.012(18)$ & n/a \\
         \hline
    \end{tabular}
    \caption{Comparison of our strange quark results with previous calculations of the same quantities. Results labeled with $\star$ are from single lattice spacing.}
    \label{tab:compare_strange}
\end{table}

\small\begin{table}
    \centering
    \begin{tabular}{|c|c|c|c|c|}
    \hline
         charm &  $g^{c,R}_S$ & $g^{c,R}_A$ &$g^{c,R}_T$ & $\langle x\rangle_c$  \\
         \hline
          \!This work (preliminary)\! & $0.015(34)$ & $-0.0026(18)$ & $0.0004(13)$ & $0.070(25)$  \\
         \hline
         $\star$ETMC'19 Nf=4 &$0.075(17)$& $-0.0098(34)$ & $-0.00024(16)$ &$0.019(9)$ \\
         \hline
         $\star$ETMC'19 Nf=2 &$0.059(18)$& $-0.0065(51)$ & $-0.0060(37)$ & n/a \\
         \hline
    \end{tabular}
    \caption{Comparison of our charm quark results with previous calculations of the same quantities. Results labeled with $\star$ are from single lattice spacing.}
    \label{tab:compare_charm}
\end{table}
\section{Conclusion}
In this work, we present the calculation of strange and charm contribution to nucleon charges and momentum fraction. We computed the disconnected diagrams on 6 lattice ensembles covering 4 lattice spacings and 3 pion masses that allow us to extrapolate to the physical point. The renormalization constants are calculated in the RI-sMOM scheme and then matched to $\overline{MS}$ scheme and evolved to $\mu=2$ GeV. We take into account the mixing for scalar charge between strange and iso-scalar, but have not included the mixing for momentum fractions. Our results on axial charge and tensor charge are consistent with previous lattice results. The scalar charge calculation needs increased statistics on finer lattices. The calculation of momentum fraction also needs more data and the evaluation of mixing with gluon operator and other flavors.


\bibliographystyle{JHEP}
\bibliography{ref}
\end{document}